\documentclass[preprint,12pt]{elsarticle}

\usepackage[ansinew]{inputenc}
\usepackage[english]{babel}
\usepackage{ifthen,shortvrb}
\usepackage[fleqn]{amsmath}
\usepackage{amsxtra,amsfonts,latexsym,amssymb,euscript}
\usepackage{amstext}
\usepackage{graphicx}
\usepackage{dcolumn}
\usepackage{subfigure}
\usepackage{natbib}
\usepackage{mathrsfs}
\usepackage{algorithm2e}

\newcommand{\IM}{\mathcal{I}_m}
\newcommand{\IN}{\mathcal{I}_n}

\newcommand{\ba}[1]{\begin{array}{#1}}
\newcommand{\ea}{\end{array}}

\begin{document}

\begin{frontmatter}
\journal{CMAME}
\title{Node-to-segment and node-to-surface interface finite elements for fracture mechanics}
\author[paggi]{M. Paggi\corref{cor}}\ead{marco.paggi@imtlucca.it}
\author[wriggers]{P. Wriggers}\ead{wriggers@ikm.uni-hannover.de}
\cortext[cor]{Corresponding author. Tel: +39-0583-4326-604, Fax:
+39-0583-4326-565.}
\address[paggi]{IMT Institute for Advanced Studies Lucca, Piazza San Francesco 19, 55100 Lucca, Italy}
\address[wriggers]{Leibniz Universi\"{a}t Hannover, Institut f\"{u}r Kontinuumsmechanik, Appelstra{\ss}e 11, 30167 Hannover, Germany}

\begin{abstract}
The topologies of existing interface elements used to discretize
cohesive cracks are such that they can be used to compute the
relative displacements (displacement discontinuities) of two
opposing segments (in 2D) or of two opposing facets (in 3D)
belonging to the opposite crack faces and enforce the cohesive
traction-separation relation. In the present work we propose a novel
type of interface element for fracture mechanics sharing some
analogies with the node-to-segment (in 2D) and with the
node-to-surface (in 3D) contact elements. The displacement gap of a
node belonging to the finite element discretization of one crack
face with respect to its projected point on the opposite face is
used to determine the cohesive tractions, the residual vector and
its consistent linearization for an implicit solution scheme. The
following advantages with respect to classical interface finite
elements are demonstrated: $(i)$ non-matching finite element
discretizations of the opposite crack faces is possible; $(ii)$ easy
modelling of cohesive cracks with non-propagating crack tips;
$(iii)$ the internal rotational equilibrium of the interface element
is assured. Detailed examples are provided to show the usefulness of
the proposed approach in nonlinear fracture mechanics problems.

\vspace{1em}
\noindent Notice: this is the authors version of a work
that was accepted for publication in Computer Methods in Applied
Mechanics and Engineering. Changes resulting from the publishing
process, such as editing, structural formatting, and other quality
control mechanisms may not be reflected in this document. A
definitive version was published in Computer Methods in Applied
Mechanics and Engineering, Vol. 300, 540--560,
DOI:10.1016/j.cma.2015.11.023
\end{abstract}
\end{frontmatter}

\section{Introduction}

The cohesive zone model (CZM) is a powerful theoretical tool to
characterize the constitutive response of cracks and study fracture
phenomena taking place across different length scales. Admitting a
continuity of tractions across the interfacial zone, displacement
discontinuities (also called relative displacements, or gaps) are
allowed to simulate material separation. Cohesive tractions acting
opposite to the relative displacements are nonlinear functions of
the gaps. Different expressions have been proposed in the literature
depending on the material, see e.g. some notable examples in
\cite{giB,aH,AC,EGGP,TV}.

Among the various applications, cohesive interfaces can be applied to
simulate the response of adhesives in composites in statics and
dynamics \cite{allix,camanho,corrado,WH}, as well as their
resistance to peeling \cite{reinoso}. In materials science, CZMs can
be efficiently used to investigate the phenomenon of intergranular
crack growth in polycrystalline materials \cite{PW11,Petal,PW12}.

From the numerical point of view, interface elements represent the
standard method to implement a cohesive crack into the finite
element method \cite{deborst,ortiz}. Considering linear
interpolation schemes, interface elements in 2D are defined in terms
of two segments coinciding with the sides of the finite elements
used to discretize the continuum on the opposite crack faces.
Analogously, in 3D, interface elements are defined in terms of two
facets. The relative opening and sliding displacements are computed
at the integration points by interpolating the nodal values. Then,
the cohesive tractions are determined according to the CZM
formulation. The integration of the contribution of the interface
element based on  the Principle of Virtual Work leads to the
residual force vector. Finally, its consistent linearization
provides the tangent stiffness matrix of the interface element. A
generalization of the basic formulation to deal with coupled
thermo-elastic problems has been proposed in \cite{HS,OBG,SP,FBK13}.
Formulations for large displacements are also available in
\cite{reinoso,ortiz,BSG,QCA01}.

\begin{figure}[h!]
\centering
\includegraphics[width=.8\textwidth,angle=0]{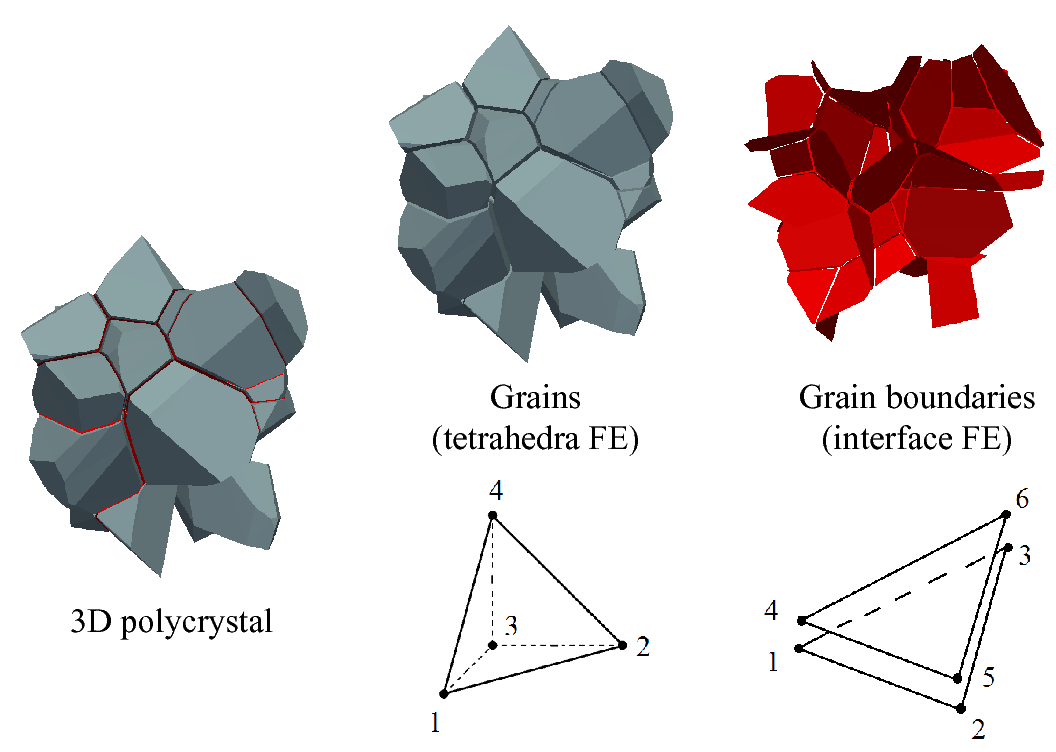}
\caption{Matching of nodes along grain boundaries requires node
duplication and a suitable data structure.}\label{fig1}
\end{figure}

\begin{figure}[h!]
\centering
\includegraphics[width=.4\textwidth,angle=0]{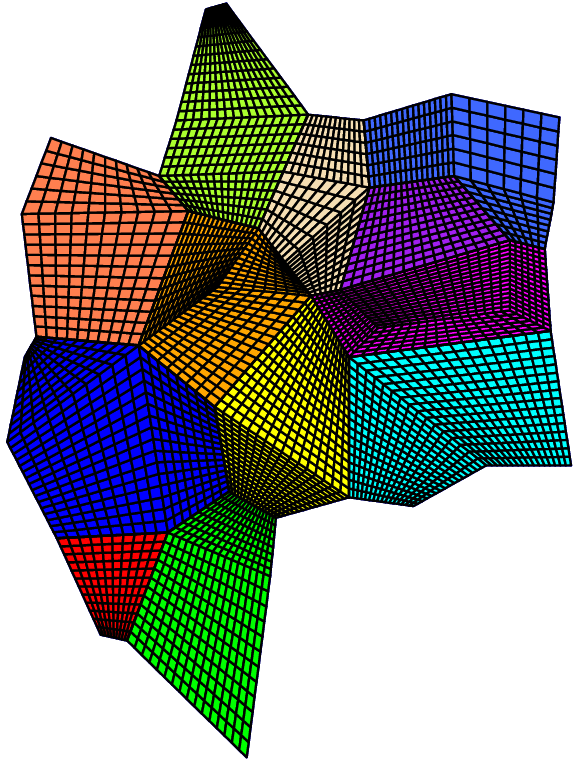}
\caption{The constraint of matching nodes at grain boundaries can
lead to non-uniform finite element meshes.}\label{fig2}
\end{figure}

Standard interface finite elements require matching of nodes at an
interface, which can be a significant constraint in many
applications. In particular, this imposes a constraint on the finite
element discretization of the domains sharing the interface that
cannot be meshed separately. Therefore, the generation of finite
element meshes of material domains separated by interfaces has to be
initially performed by considering them as perfectly bonded
together. As the next step, which is not usually possible to be done
in commercial mesh generation software, the nodes belonging to the
internal boundaries are duplicated and the new interface elements
are assembled by specifying their connectivity matrix
\cite{PW11,PW12}. This procedure requires a complex data management
in 3D geometries as in polycrystalline materials (see
Fig.\ref{fig1}), where all the nodes belonging to grain boundary
facets have to be identified and stored in a suitable data format
for their duplication \cite{Petal}. This constraint can induce a non
uniform mesh discretization of the grains, as it can be already seen
in 2D problems as in Fig.\ref{fig2}, where the block command of FEAP
\cite{FEAP} was used to generate a structured finite element mesh of
the continuum \cite{Petal}. Therefore, to overcome this problem,
unstructured meshes with triangular or tetrahedra elements are
usually preferred.

\begin{figure}[h!]
\centering
\includegraphics[width=.45\textwidth,angle=0]{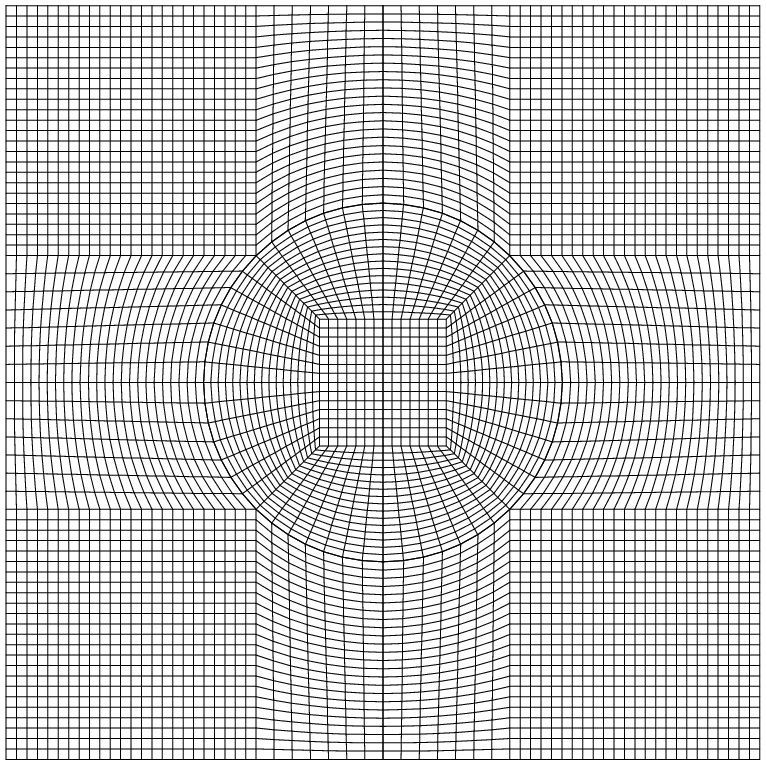}
\caption{Fiber-reinforced metal-matrix composites: the fiber behaves
almost as a rigid body and it could be discretized with a coarser
finite element mesh than that used for the matrix.}\label{fig3}
\end{figure}

\begin{figure}[h!]
\centering
\includegraphics[width=.55\textwidth,angle=0]{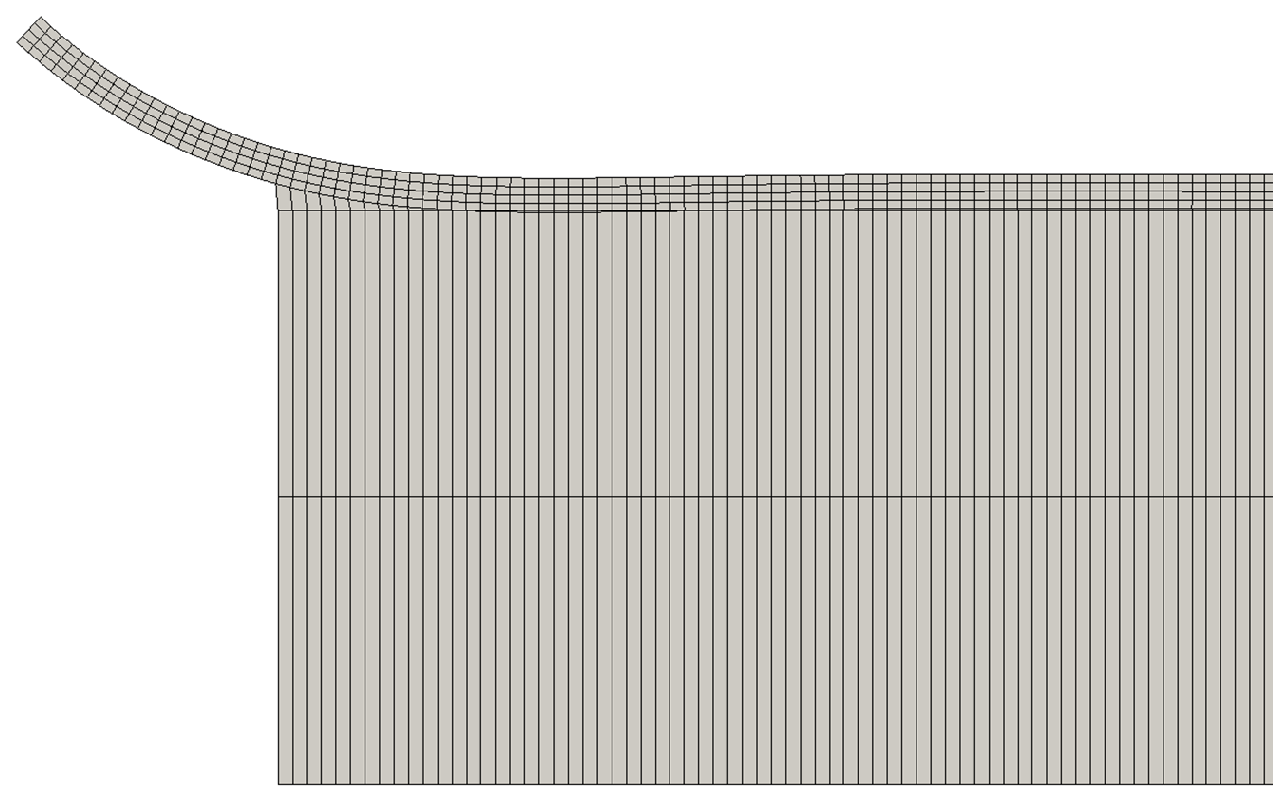}
\caption{Peeling of a deformable layer from a stiff substrate: the
substrate could be meshed with a coarser finite element mesh than
that used for the layer.}\label{fig4}
\end{figure}

In composites, on the other hand, it is frequent to have material
components with very different elastic properties. This is for
instance the case of Aluminum metal-matrix composites reinforced by
stiff ceramic fibers (Fig.\ref{fig3}), or in the case of deformable
polymeric layers bonded to a very stiff substrate, as for epoxy
layers bonded on glass used in photovoltaic applications
(Fig.\ref{fig4}). The stiffer component of the composite system
behaves almost as a rigid body with negligible deformation.
Therefore, it would be computationally efficient to adopt a coarse
finite element discretization for the rigid component, albeit
keeping fine the finite element discretization of the interface and
of the deformable material.

In all of these problems, a much simpler and computationally more
efficient mesh generation procedure would be to mesh each material
region independently from the others and then insert interface
elements along their internal boundaries capable to deal with non
matching nodes. In the case of mesh adaptivity based on error
control during nonlinear fracture mechanics simulations, the
possibility of dealing with different mesh dicretizations of the
continuum domains would also be very useful.

In the present study, we propose two new interface finite elements
for interface fracture problems inspired by contact mechanics able
to deal with non matching nodes: one for 2D applications and called
\emph{node-to-segment interface element}, and another for 3D
simulations and called \emph{node-to-surface interface element}.
Their mathematical formulation and the matrix form for the finite
element implementation are detailed in Sections 2 and 3. Section 4
presents the mesh generation algorithms to be used in conjunction
with the new interface elements. Patch tests are discussed in
Section 5 to show that the new elements are able to transfer a
uniform stress state in the case of uniform loading. Simple but
relevant numerical applications showing the capabilities of the new
finite element formulations are finally discussed in Section 6.
Conclusions and perspectives complete the article.

\section{Continuum mechanics framework}

\begin{figure}[h!]
\centering
\includegraphics[width=.6\textwidth,angle=0]{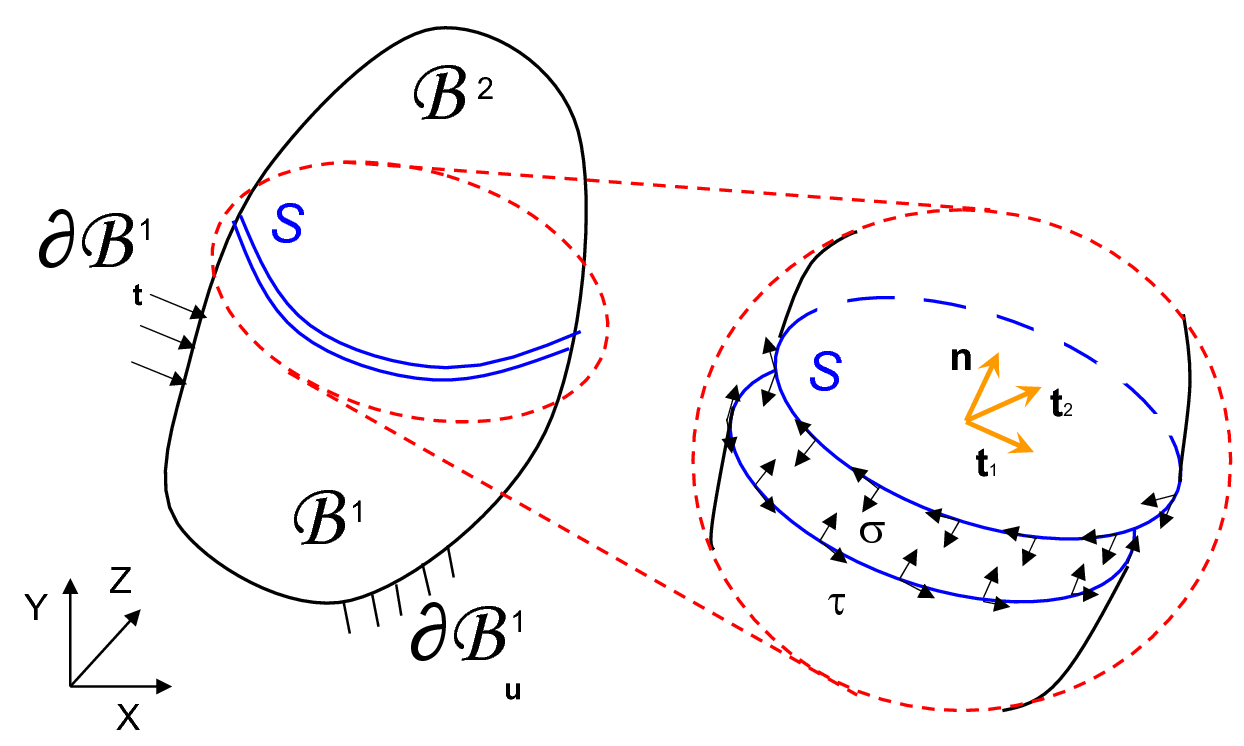}
\caption{A sketch of two continuum bodies separated by a cohesive
interface.}\label{fig5}
\end{figure}

In the small deformation setting, let us consider a pair of
deformable bodies in the reference undeformed configuration
$\mathscr{B}^{1}$, $\mathscr{B}^{2}$ $\subset \mathbb{R}^{n_{dim}}$,
where $n_{dim}$ is equal to 2 or 3 for 2D or 3D problems,
respectively. In general, both bodies are subject to volume forces
$\mathbf{F}_{v}^{i}$ $(i=1,2)$, and to boundary conditions under the
form of tractions, $\mathbf{t}^{i}=\hat{\mathbf{t}}^{i}$ $(i=1,2)$
on $\partial \mathscr{B}_{\mathbf{t}}^{i}$, and displacements,
$\mathbf{u}^{i}=\hat{\mathbf{u}}^{i}$ $(i=1,2)$ on $\partial
\mathscr{B}_{\mathbf{u}}^{i}$. The bodies might have different
linear or nonlinear constitutive relations and material properties
(see Fig.\ref{fig5}).

The interface between the deformable bodies is a region of lower
spatial dimension denoted by $S \subset \mathbb{R}^{n_{dim} - 1}$
where we allow displacement discontinuities, see Fig.\ref{fig5}.
Along the interface, defined by two opposing segments in 2D or
facets in 3D, a local reference system can be introduced in relation
to the modes of deformation of fracture mechanics, i.e., opening (or
Mode I), sliding (or Mode II), and tearing (or Mode III).

Therefore, a normal vector $\mathbf{n}$ and a tangent vector
$\mathbf{t}$ can be introduced along a 2D interface segment.
Similarly, a normal vector $\mathbf{n}$ and two tangent vectors
$\mathbf{t}_1$ and $\mathbf{t}_2$ can be defined along a 3D
interface surface (see Fig.\ref{fig5}). Since we are in the
hypothesis of a zero-thickness interface and we are restricting our
study to the small deformation setting, these vectors can be
computed from the coordinates $\mathbf{x}$ of the nodes belonging to
any of the two opposing segments or facets. Cohesive tractions
opposing to the relative displacements can be interpreted in the
framework of configurational forces \cite{CF}, since their intensity
and direction depend on the relative motion of the two bodies
sharing the interface.

The virtual work of the interface tractions contribute to the
Principle of Virtual Work of the whole system.  This is the starting
point to derive the finite element formulation. We recall that, for
a standard interface element, we have:
\begin{equation}\label{WF}
\delta\Pi_{\mathrm{intf}}
 (\mathbf{u}, \delta \mathbf{u}) = \delta\mathbf{u}^{\mathrm{T}} \left[ \int_{S_0}
\left(\dfrac{\partial\mathbf{g}_{\mathrm{loc}}}{\partial\mathbf{u}}\right)^{\mathrm{T}}\,\mathbf{T}\mathrm{d}S\right],
\hspace{0.2cm} \forall \delta \mathbf{u} \in \mathscr{V}\text{,}
\end{equation}
where $\mathbf{g}_{\mathrm{loc}}$ is the gap vector that accounts
for opening, sliding, and tearing displacements between the two
sides of the interface and it represents the work conjugated
magnitude to the cohesive tractions $\mathbf{T}$. The variable
$\delta \mathbf{u} \in \mathscr{V}$ denotes the vector of the
kinematically admissible virtual displacements.

\section{Node-to-segment interface finite element}

\begin{figure}[h!]
\centering \subfigure[Traditional interface
element]{\includegraphics[width=.6\textwidth,angle=0]{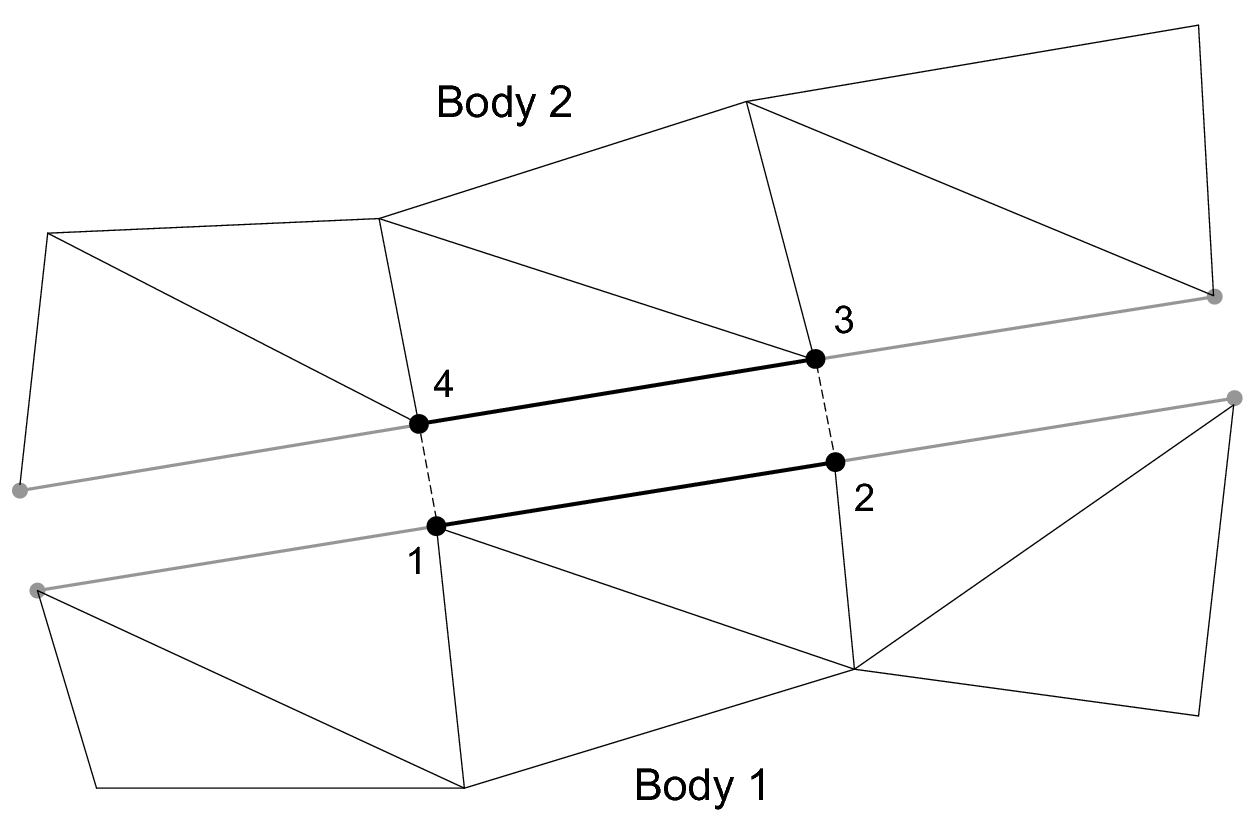}\label{fig6a}}\\
\subfigure[Node-to-segment interface
element]{\includegraphics[width=.6\textwidth,angle=0]{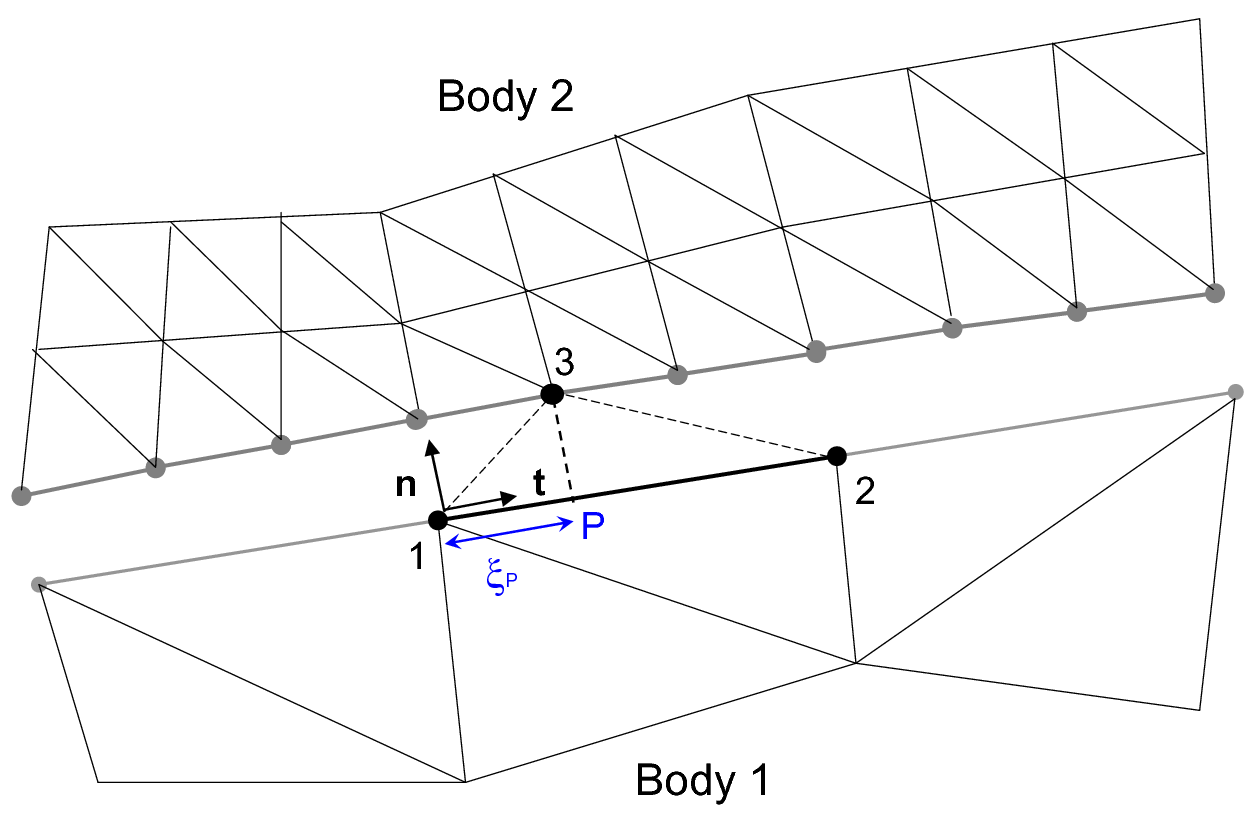}\label{fig6b}}
\caption{(a) Classical four-node interface element. (b) Proposed
node-to-segment interface element.}\label{fig6}
\end{figure}

After introducing the finite element discretization of the two
bodies sharing an interface by using linear triangular or
quadrilateral elements, standard interface elements are assembled by
pairing two opposing segments belonging to body 1 and 2, defined in
terms of four nodes (see Fig.\ref{fig6a}). Clearly, the use of these
interface elements requires matching of the nodes of body 1 and body
2 at the common interface. \emph{At present, there is no method
available in the literature to deal with non-conforming cohesive
interfaces.}

To avoid this constraint, let's define a new interface element
relating the displacements of one node belonging to the boundary of
the body 2 to the displacements of its projection normal to the
corresponding segment belonging to the body 1, see Fig.\ref{fig6b}.
We have therefore a topology analogous to a node-to-segment
formulation used in contact mechanics \cite{contact}, with the
simplification that, in fracture mechanics applications under small
displacements, pairing between the segment and the node has not to
be updated during the simulation and it can be defined at the mesh
generation stage.

For this interface element, the contribution to the Principle of
Virtual Work reads:
\begin{equation}\label{WF2D}
\delta\Pi_{\mathrm{intf}}
 (\mathbf{u}, \delta \mathbf{u}) = \delta\mathbf{u}^{\mathrm{T}}
\left(\dfrac{\partial\mathbf{g}_{\mathrm{loc},P}}{\partial\mathbf{u}}\right)^{\mathrm{T}}\,\mathbf{T}_P\,l_{el},
\hspace{0.2cm} \forall \delta \mathbf{u} \in \mathscr{V}
\end{equation}
where $\mathbf{g}_{\mathrm{loc},P}$ is the gap vector in the local
reference system computed from the relative displacements of node 3
and of its projection $P$ on the segment $1-2$, and $\mathbf{T}_P$
is the corresponding traction vector function of the gaps according
to the selected CZM. These quantities are multiplied by the length
$l_{el}$, which is the area of influence of the node $3$ and is
related to the finite element discretization of the interface from
the side of body 2. The present formulation is valid for both
structured and unstructured finite element meshes. More
specifically, for each node belonging to the side of the interface
having the finer discretization, which is the third node of each new
interface element to be constructed according to the sketch in
Fig.\ref{fig6b}, $l_{el}$ is computed as the sum of half the
distances between that node and its closest neighbors. For edge
nodes, only one neighbor has to be considered in the computation of
$l_{el}$.

After introducing the finite element discretization, the continuum
displacement vector $\mathbf{u}$ can be replaced by the nodal
displacement vector
$\mathbf{d}=(u_1,v_1,u_2,v_2,u_3,v_3)^{\mathrm{T}}$ in
Eq.\eqref{WF2D}:
\begin{equation}\label{WF2Dd}
\delta\Pi_{\mathrm{intf}}
 (\mathbf{d}, \delta \mathbf{d}) = \delta\mathbf{d}^{\mathrm{T}}
\left(\dfrac{\partial\mathbf{g}_{\mathrm{loc},P}}{\partial\mathbf{d}}\right)^{\mathrm{T}}\,\mathbf{T}_P\,l_{el}
\end{equation}
This provides the expression of the element residual vector
$\mathbf{f}^e$:
\begin{equation}
\mathbf{f}^e
=\left(\dfrac{\partial\mathbf{g}_{\mathrm{loc},P}}{\partial\mathbf{d}}\right)^{\mathrm{T}}\,\mathbf{T}_P\,l_{el}
\end{equation}
The local gap vector in the point $P$,
$\mathbf{g}_{\mathrm{loc},P}=(g_n,g_t)_{P}^{\mathrm{T}}$, can be
determined from the nodal displacement vector as follows:
\begin{equation}\label{gap2d}
\mathbf{g}_{\mathrm{loc},P}=\mathbf{R}\mathbf{B}_P\mathbf{d},
\end{equation}
where $\mathbf{R}$ is the rotation matrix defined in terms of the
unit vectors $\mathbf{n}=(n_x, n_y)^{\mathrm{T}}$ and
$\mathbf{t}=(t_x, t_y)^{\mathrm{T}}$ that are related to the
coordinates of the nodes 1 and 2, see Appendix for their explicit
expressions:
\begin{equation}
\mathbf{R}=\left[
              \begin{array}{cc}
                n_x & n_y \\
                t_x & t_y \\
              \end{array}
            \right]
\end{equation}

The operator $\mathbf{B}_P$ is defined as:
\begin{equation}\label{B2D}
\mathbf{B}_P=\left[
              \begin{array}{cccccc}
                -N_1 & 0 & -N_2 & 0 & 1 & 0\\
                 0 & -N_1 & 0 & -N_2 & 0 & 1\\
              \end{array}
            \right]_P
\end{equation}
where $N_1=\xi/l$ and $N_2=(1-\xi/l)$ are the linear shape functions
for the nodes $1$ and $2$, dependent on the surface coordinate $\xi$
as in Fig.\ref{fig6b}. The length $l$ is defined by
$l=\|\mathbf{x}_2-\mathbf{x}_1\|$. The shape functions have to be
computed in correspondence to the point $P$, i.e., for
$\xi_P=(\mathbf{x}_3-\mathbf{x}_1)\cdot \mathbf{t}$.

Introducing Eq.\eqref{gap2d} into Eq.\eqref{WF2Dd}, the final matrix
form for the element residual vector is derived:
\begin{equation}\label{residual2D}
\mathbf{f}^e
=\mathbf{B}_P^{\mathrm{T}}\,\mathbf{R}^{\mathrm{T}}\mathbf{T}_P\,l_{el}
\end{equation}

The element tangent stiffness matrix for an implicit solution scheme
is computed by performing a consistent linearization of the
residual:
\begin{equation}\label{stif2D}
\mathbf{K}^e=\dfrac{\partial \mathbf{f}^e}{\partial \mathbf{d}}=
\mathbf{B}_P^{\mathrm{T}}\,\mathbf{R}^{\mathrm{T}}\dfrac{\partial
\mathbf{T}_P}{\partial \mathbf{d}}\,l_{el}
\end{equation}
where $\partial \mathbf{T}_P/\partial \mathbf{d}$ is obtained by a
chain rule differentiation:
\begin{equation}
\dfrac{\partial \mathbf{T}_P}{\partial \mathbf{d}}=\dfrac{\partial
\mathbf{T}_P}{\partial \mathbf{g}_{\mathrm{loc},P}} \dfrac{\partial
\mathbf{g}_{\mathrm{loc},P}}{\partial
\mathbf{d}}=\mathbf{C}_P\mathbf{R}\mathbf{B}_P
\end{equation}
The tangent constitutive matrix $\mathbf{C}_P$ depends on the form
of the cohesive traction-separation relation
$\mathbf{T}=(\sigma,\tau)^{\mathrm{T}}$ vs.
$\mathbf{g}_{\mathrm{loc}}=(g_n,g_t)^{\mathrm{T}}$, i.e., on the CZM
expression. Its symbolic form reads:
\begin{equation}
\mathbf{C}_P=\left[
              \begin{array}{cccc}
                \dfrac{\partial\sigma}{\partial g_n} & \dfrac{\partial\sigma}{\partial g_t}\\
                \dfrac{\partial\tau}{\partial g_n}   & \dfrac{\partial\tau}{\partial g_t}\\
              \end{array}
            \right]_P
\end{equation}

It has to be remarked that the proposed node-to-segment interface
element satisfies the internal rotational equilibrium, while the
standard 4-nodes interface element does not, as pointed out in
\cite{geers}.

\section{Node-to-surface interface finite element}

\begin{figure}[h!]
\centering \subfigure[Traditional interface
element]{\includegraphics[width=.6\textwidth,angle=0]{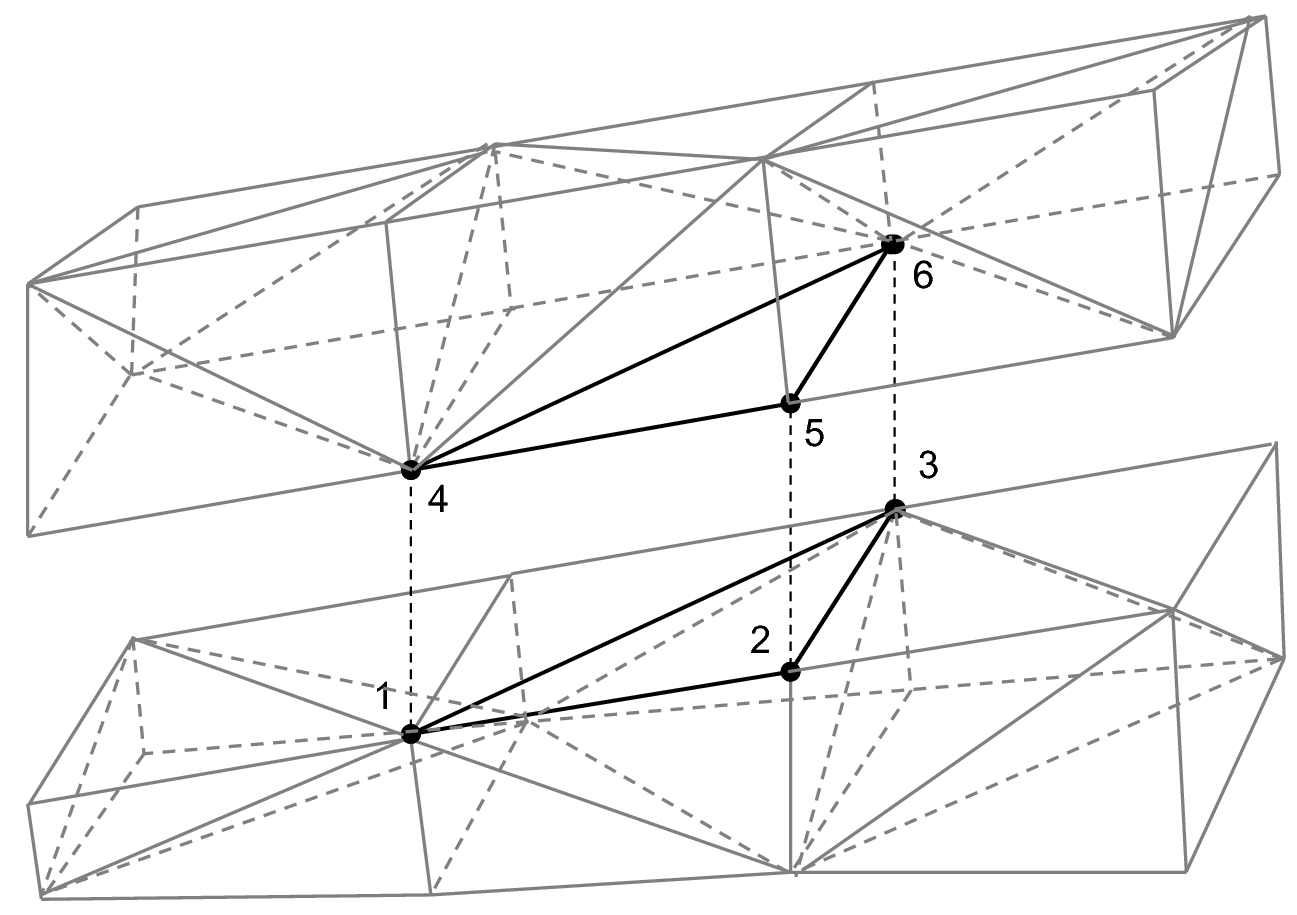}\label{fig7a}}\\
\subfigure[Node-to-surface interface
element]{\includegraphics[width=.6\textwidth,angle=0]{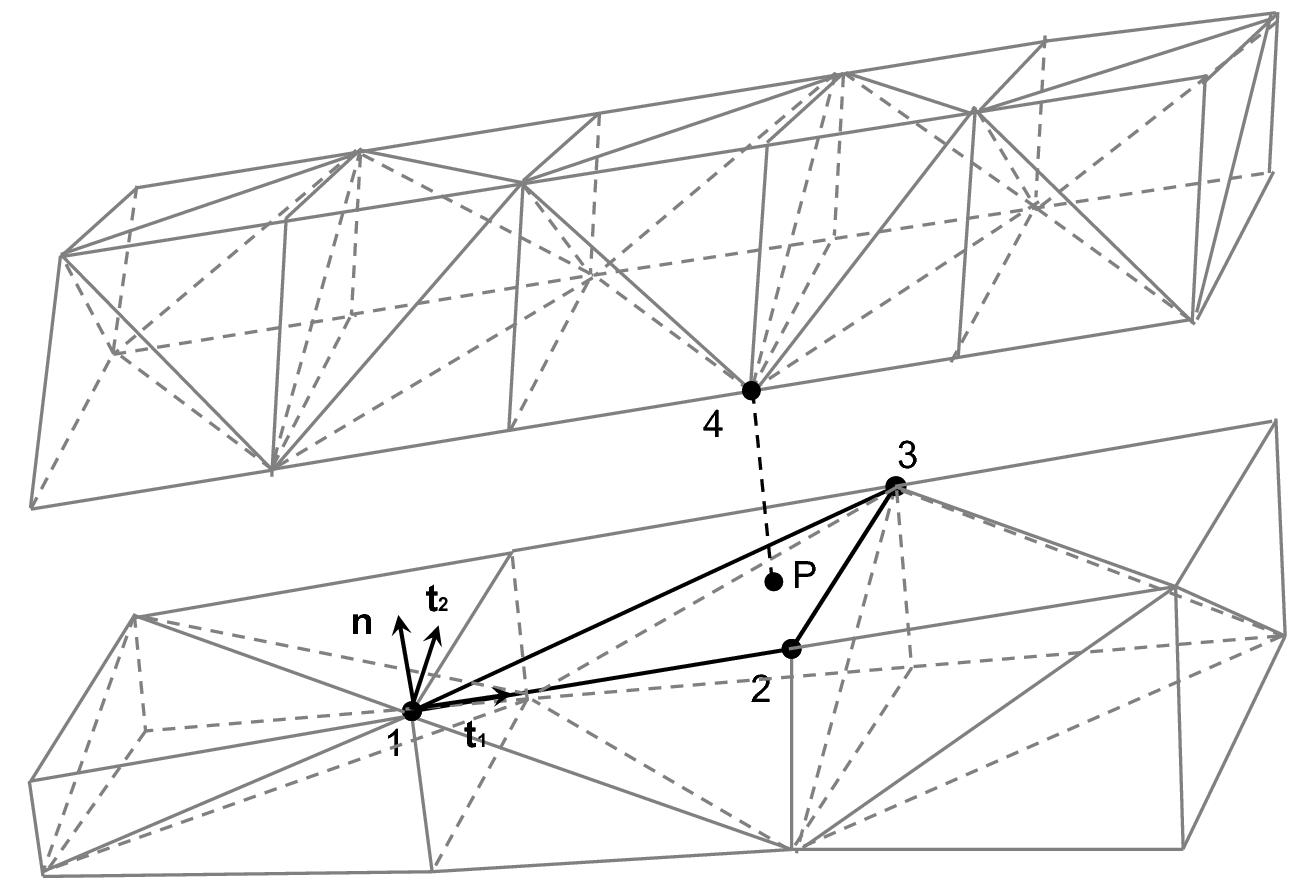}\label{fig7b}}
\caption{(a) classical six-node interface element. (b) Proposed
node-to-surface interface element.}\label{fig7}
\end{figure}

For 3D problems, like for intergranular crack propagation in
polycrystalline materials studied in \cite{Petal}, the finite
element discretization of the bulk is usually performed by using
tetrahedra finite elements, due to the high versatility in meshing
complex polyhedral domains. In case of linear tetrahedra, standard
interface elements are linking the edges of two tetrahedra and are
represented by two triangles in 3D with 3 nodes each. A sketch of
this standard interface element is shown in Fig.\ref{fig7a}.

For different structured finite element discretizations of the two
bodies, it is convenient to introduce a new interface element that
relates the displacements of one node belonging to the boundary of
body 2 to the displacements of its projection normal to the
corresponding triangular facet belonging to body 1, see
Fig.\ref{fig7b}. We have therefore a topology analogous to a
node-to-surface formulation used in contact mechanics
\cite{contact}, again with fixed pairing defined during mesh
generation due to the small displacement scenario.

For this interface element, the contribution to the Principle of
Virtual Work reads:
\begin{equation}\label{WF3D}
\delta\Pi_{\mathrm{intf}}
 (\mathbf{u}, \delta \mathbf{u}) = \delta\mathbf{u}^{\mathrm{T}}
\left(\dfrac{\partial\mathbf{g}_{\mathrm{loc},P}}{\partial\mathbf{u}}\right)^{\mathrm{T}}\,\mathbf{T}_P\,A_{el}k,
\quad \forall \delta \mathbf{u} \in \mathscr{V}
\end{equation}
where $\mathbf{g}_{\mathrm{loc},P}$ is related to the relative
displacements of node 4 and of its projection $P$ onto the facet
$1-2-3$. $\mathbf{T}_P$ is the corresponding traction vector given
by the CZM relation. For each interface element, these quantities
are multiplied by the area $A_{el}$ and by the coefficient $k$. For
structured interface discretizations, $A_{el}$ is the area of any of
the $e$ facets converging to the node 4 on the side of the interface
having the finer discretization. The coefficient $k$ can be computed
as the ratio between the number $e$ of facets converging to the node
4, divided by three times the number $q$ of facets belonging to the
opposite side of the interface (having the coarser discretization)
with minimum distance from the node 4. The number $q$ is typically
equal to 1 if the normal projection of node 4 to the opposite side
of the interface is inside a facet, or is equal to 2 if it falls on
the edge between two adjoining facets. Edge nodes are treated in the
same way as the internal nodes. A practical example showing the
various possible cases is illustrated in the next section which
treats the algorithms for mesh generation in the 2D and in the 3D
cases.

After introducing the finite element discretization, we can replace
the continuum displacement vector $\mathbf{u}$ with the nodal
displacement vector
$\mathbf{d}=(u_1,v_1,w_1,u_2,v_2,w_2,u_3,v_3,w_3)^{\mathrm{T}}$ in
Eq.\eqref{WF3D}:
\begin{equation}\label{WF3Dd}
\delta\Pi_{\mathrm{intf}}
 (\mathbf{d}, \delta \mathbf{d}) = \delta\mathbf{d}^{\mathrm{T}}
\left(\dfrac{\partial\mathbf{g}_{\mathrm{loc},P}}{\partial\mathbf{d}}\right)^{\mathrm{T}}\,\mathbf{T}_P\,A_{el}k
\end{equation}
which provides the expression of the element residual vector
$\mathbf{f}^e$:
\begin{equation}\label{residual3D}
\mathbf{f}^e
=\left(\dfrac{\partial\mathbf{g}_{\mathrm{loc},P}}{\partial\mathbf{d}}\right)^{\mathrm{T}}\,\mathbf{T}_P\,A_{el}k
\end{equation}
The local gap vector in the point $P$,
$\mathbf{g}_{\mathrm{loc},P}=(g_n,g_{t1},g_{t2})_{P}^{\mathrm{T}}$,
can be determined from the nodal displacement vector, see
Eq.\eqref{gap2d}, where $\mathbf{R}$ is now a rotation matrix
defined in terms of the unit vectors $\mathbf{n}=(n_x, n_y,
n_z)^{\mathrm{T}}$, $\mathbf{t}_1=(t_{1x},
t_{1y},t_{1z})^{\mathrm{T}}$, and $\mathbf{t}_2=(t_{2x},
t_{2y},t_{2z})^{\mathrm{T}}$, see Fig.\ref{fig7b}:
\begin{equation}
\mathbf{R}=\left[
              \begin{array}{ccc}
                n_x & n_y & n_z \\
                t_{1x} & t_{1y} & t_{1z} \\
                t_{2x} & t_{2y} & t_{2z} \\
              \end{array}
            \right]
\end{equation}
The components of $\mathbf{R}$ can be related to the coordinates of
the nodes 1, 2, and 3, $\mathbf{x}_i=(x_i,y_i,z_i)^{\mathrm{T}}$
$(i=1,2,3)$, see the expressions in the Appendix.

The operator $\mathbf{B}_P$ is defined as:
\begin{equation}\label{B3D}
\mathbf{B}_P=\left[
              \begin{array}{cccccccccccc}
                -N_1 & 0 & 0& -N_2 & 0 & 0& -N_3 & 0 & 0& 1 & 0 & 0 \\
                   0 & -N_1 & 0& 0 & -N_2 & 0& 0& -N_3 & 0& 0 & 1 & 0 \\
                   0 & 0 & -N_1 & 0 & 0 & -N_2 & 0 & 0 & -N_3 & 0 & 0 & 1\\
              \end{array}
            \right]_P
\end{equation}
where linear shape functions $N_1$, $N_2$, and $N_3$ are introduced
to interpolate the nodal quantities over the $1-2-3$ plane. Denoting
with $\mathbf{x}^*=(\xi,\eta,s)^{\mathrm{T}}$ the coordinates in the
local reference system defined by the local frame $\mathbf{n}$,
$\mathbf{t}_1$ and $\mathbf{t}_2$ with origin in the node $1$,
obtained by pre-multiplying the vectors $(\mathbf{x}-\mathbf{x}_1)$
by the rotation matrix $\mathbf{R}$,
$\mathbf{x}^*=\mathbf{R}(\mathbf{x}-\mathbf{x}_1)$, we can write
\begin{subequations}
\begin{align}
N_1&=a_1\xi+b_1\eta+c_1\\
N_2&=a_2\xi+b_2\eta+c_2\\
N_3&=a_3\xi+b_3\eta+c_3
\end{align}
\end{subequations}
with coefficients $a_i$, $b_i$, and $c_i$ $(i=1,2,3)$:
\begin{subequations}
\begin{align*}
a_1&=\dfrac{\eta_2-\eta_3}{\xi_2\eta_3-\eta_2\xi_3},\quad
b_1=\dfrac{\xi_3-\xi_2}{\xi_2\eta_3-\eta_2\xi_3},\quad
c_1=1\\
a_2&=\dfrac{\eta_3}{\xi_2\eta_3-\eta_2\xi_3},\quad
b_2=-\dfrac{\xi_3}{\xi_2\eta_3-\eta_2\xi_3},\quad
c_2=0\\
a_3&=-\dfrac{\eta_2}{\xi_2\eta_3-\eta_2\xi_3},\quad
b_3=\dfrac{\xi_2}{\xi_2\eta_3-\eta_2\xi_3},\quad
c_3=0
\end{align*}
\end{subequations}
Note that the shape functions entering the matrix $\mathbf{B}_P$ in
Eq.\eqref{B3D} have to be computed with respect to the coordinates
of the point $P$, i.e., for $\xi=\xi_4$ and $\eta=\eta_4$.

Introducing the expression for the gap into Eq.\eqref{WF3Dd} we
obtain:
\begin{equation}\label{WF3Dd_fin}
\mathbf{f}^e
=\mathbf{B}_P^{\mathrm{T}}\,\mathbf{R}^{\mathrm{T}}\mathbf{T}_P\,A_{el}k
\end{equation}
The element tangent stiffness matrix is derived from the consistent
linearization of the element residual vector:
\begin{equation}\label{stif3D}
\mathbf{K}^e=\dfrac{\partial \mathbf{f}^e}{\partial \mathbf{d}}=
\mathbf{B}_P^{\mathrm{T}}\,\mathbf{R}^{\mathrm{T}}\dfrac{\partial
\mathbf{T}_P}{\partial \mathbf{d}}\,A_{el}k
\end{equation}
where $\partial \mathbf{T}_P/\partial \mathbf{d}$ is computed by a
chain rule differentiation:
\begin{equation}
\dfrac{\partial \mathbf{T}_P}{\partial \mathbf{d}}=\dfrac{\partial
\mathbf{T}_P}{\partial \mathbf{g}_{\mathrm{loc},P}} \dfrac{\partial
\mathbf{g}_{\mathrm{loc},P}}{\partial
\mathbf{d}}=\mathbf{C}_P\mathbf{R}\mathbf{B}_P
\end{equation}
The tangent constitutive matrix $\mathbf{C}_P$ depends on the form
of the cohesive traction-separation relation
$\mathbf{T}=(\sigma,\tau_1,\tau_2)^{\mathrm{T}}$ vs.
$\mathbf{g}_{\mathrm{loc}}=(g_n,g_{t1},g_{t2})^{\mathrm{T}}$, i.e.,
on the CZM expression. Its symbolic form reads:
\begin{equation}
\mathbf{C}_P=\left[
              \begin{array}{cccc}
                \dfrac{\partial\sigma}{\partial g_n} & \dfrac{\partial\sigma}{\partial g_{t1}} & \dfrac{\partial\sigma}{\partial g_{t2}}\\
                \dfrac{\partial\tau_1}{\partial g_n} & \dfrac{\partial\tau_1}{\partial g_{t1}} & \dfrac{\partial\tau_1}{\partial g_{t2}}\\
                \dfrac{\partial\tau_2}{\partial g_n} & \dfrac{\partial\tau_2}{\partial g_{t1}} & \dfrac{\partial\tau_2}{\partial g_{t2}}
              \end{array}
            \right]_P
\end{equation}

As for the node-to-segment interface element is concerned, also this
node-to-surface interface element satisfies the internal rotational
and torsional equilibrium in addition to the translational one,
while standard interface elements do not.

\section{Mesh generation procedures for the new interface elements}

Since the proposed interface elements do not require the same finite
element discretization on both sides of the interface, the
operations of mesh generation can be highly simplified with respect
to standard interface elements. Material domains can be meshed
independently from each other, with different finite element
discretizations. In the case of structured meshes with uniform mesh
size on both sides of the interface, then we can call $h_1$ and
$h_2$ the interface mesh sizes of body 1 and body 2, respectively.
This is a common situation resulting from mesh generators when the
number of divisions/element size are specified by the user for the
internal boundaries. After meshing the domains, nodes belonging to
different material regions should not be tied. We also remark that
the present node-to-segment interface element formulation can also
comply with non-structured finite element meshes.

\begin{figure}[h!]
\centering
\includegraphics[width=.6\textwidth,angle=0]{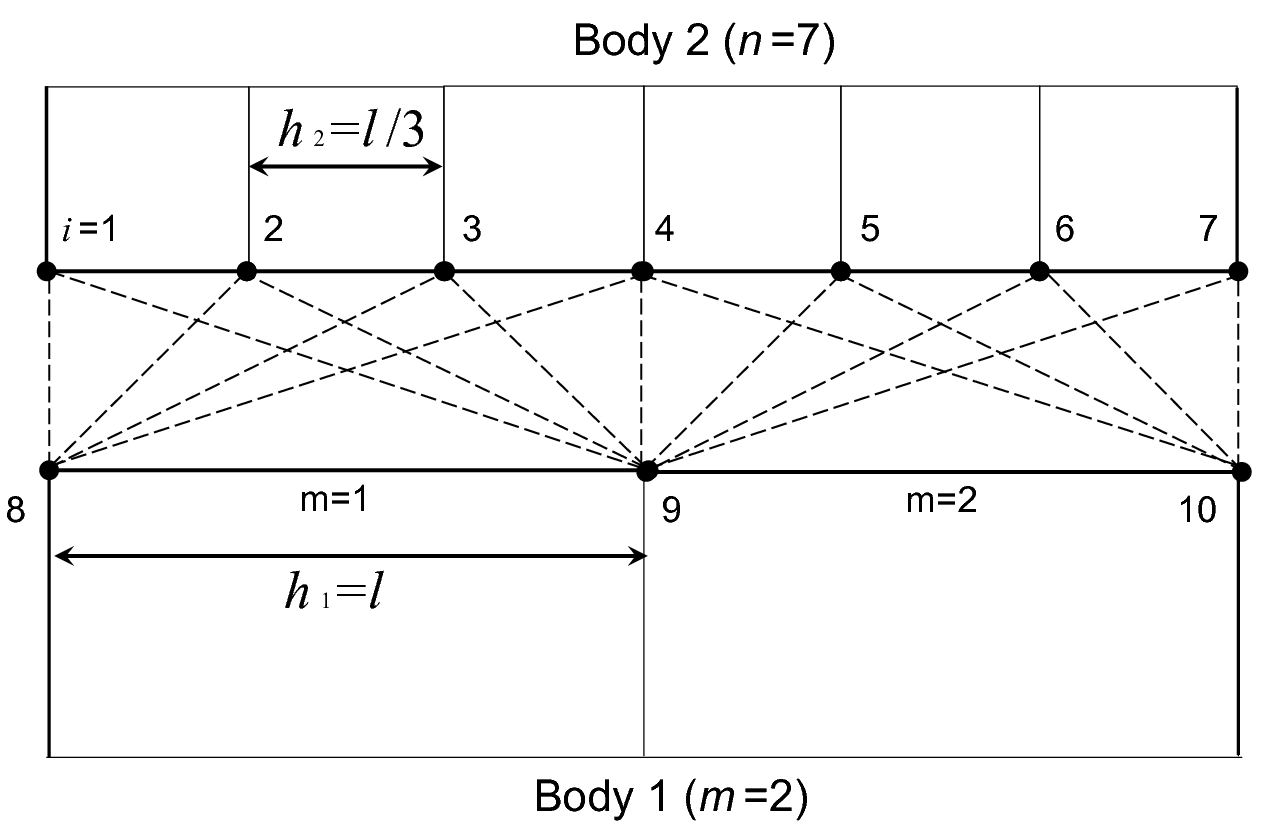}
\caption{An example showing 2D interface mesh generation with
node-to-segment interface elements.}\label{fig8}
\end{figure}

In the case of node-to-segment interface elements, see
Fig.\ref{fig8}, mesh generation can be implemented via a loop over
the nodes $i$ belonging to the set $\IN$ of the elements
discretizing the boundary of body 2. The numerosity of this set is
$\#\IN=n$ and the distance between two consecutive nodes is $l$. On
the opposite side of the interface it is possible to identify a set
of segments $\IM$ belonging to body 1, with numerosity $\#\IM=m$. In
the sketch in Fig.\ref{fig8} we have for instance $n=7$ and $m=2$
and the ratio between element sizes is $h_1/h_2=3$. For each node
$i\in \IN$, the problem is to find the segment $m\in\IM$ having the
minimum distance from $i$. The connectivity matrix of the
node-to-segment finite element to be generated will be composed by
the node numbers defining the segment, plus the number of the node
$i$. If a node has two segments with the same minimal distance, then
two interface elements have to be assembled, one for each segment.
For instance, in the example in Fig.\ref{fig8}, the node 4 has to be
paired with the segment $m=1$ (defined by the nodes 8 and 9) and
with the segment $m=2$ (defined by the nodes 9 and 10).

In the case of structured meshes, the length of influence of the
nodes, $l_{el}$, is equal to $l/3$ for $i=2,\dots,6$, and it is
equal to $l/6$ for the boundary nodes $i=1$ and $i=7$, see
Fig.\ref{fig8}. For unstructured meshes, on the other hand, $l_{el}$
varies from node to node and it is equal to the sum of half the
distances between that node and its two neighbors. For edge nodes,
only one half a distance with its neighbor has to be computed. The
operations of the 2D mesh generation procedure are summarized in
Algorithm \ref{algo2D}.

\begin{algorithm}[h!]
\caption{2D interface mesh generation algorithm} \label{algo2D}
\vspace*{.1cm}\hrule\vspace*{.1cm} ~~\textbf{Input}: set of nodes
$\IN$ belonging to the interface of body 2 $(\#\IN=n)$; set of
segments $\IM$ $(\#\IM=m)$ of the interface discretization of body
1. \vspace*{.1cm}\hrule\vspace*{.1cm}
\begin{enumerate}
\item[] \textbf{LOOP} $i=1,\dots,n$
\begin{enumerate}
\item[] Compute the length of influence of the node $i$, $l_{el}$, as the sum of half the distances between the node $i$ and its neighboring nodes. For an edge node, only one neighbor exists;
\item[] Determine the number $q$ of segments of the set $\IM$ with minimum distance
from the node $i$;
\item[] \textbf{LOOP} $j=1,\dots,q$
\item[] \quad Determine the node numbers defining the segment $j$;
\item[] \quad Construct the connectivity matrix of the interface element as a list of nodes defining the segment $j$, plus the node $i$;
\item[] \textbf{END LOOP}
\end{enumerate}
\item[] \textbf{END LOOP}
\end{enumerate}
\vspace*{.1cm}\hrule\vspace*{.1cm} ~~\textbf{Output}: Connectivity
matrix of the node-to-segment interface elements to pair different
structured or unstructured finite element meshes of the continuum.
\vspace*{.1cm}\hrule\vspace*{.1cm}
\end{algorithm}

\begin{figure}[h!]
\centering
\includegraphics[width=.55\textwidth,angle=0]{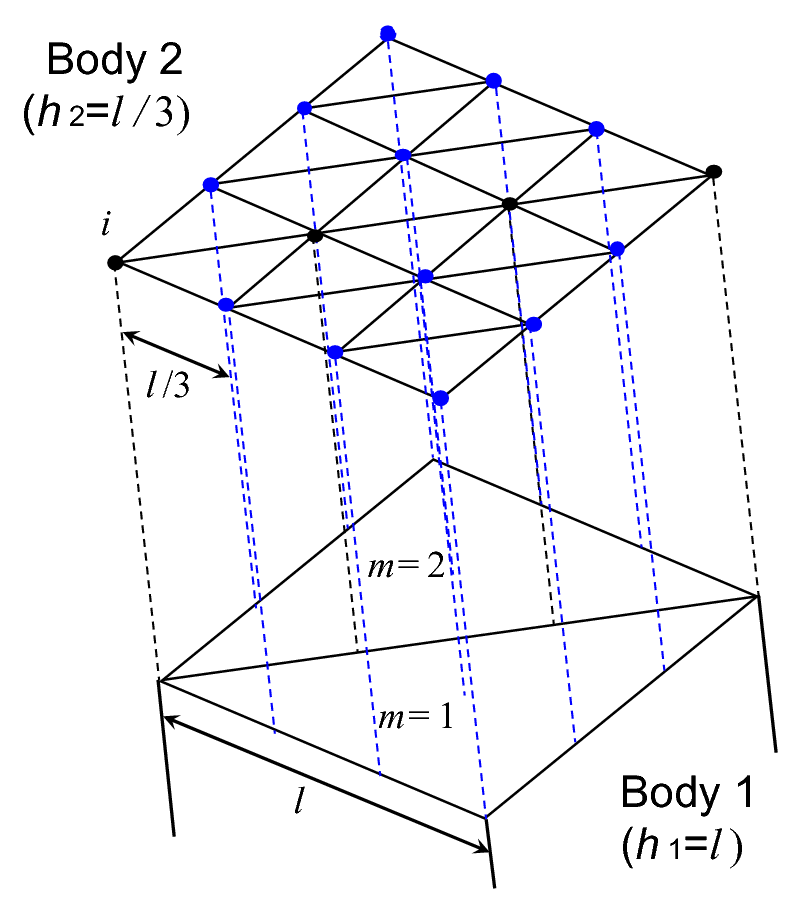}
\caption{An example showing 3D interface mesh generation with
node-to-surface interface elements: black nodes have their
orthogonal projection to the interface onto an internal edge. They
will require assembling of two interface elements, one for the facet
$m=1$, and another for the facet $m=2$.}\label{fig9}
\end{figure}

Mesh generation in 3D requires again a loop over the nodes $i\in
\IN$. We consider here for the sake of simplicity structured mesh
discretizations of the bulk realized with linear tetrahedra having a
uniform mesh size from each side of the interface, $h_1$ and $h_2$,
for body 1 and body 2, respectively. The area $A_{el}$ is related to
$h_2$ and its corresponds to the area of any facet converning to
$i$, see the example in Fig.\ref{fig9}.

For each node $i$, the facet $m$ of the finite element belonging to
the interface of body 1, with minimal distance from $i$, is
determined. In the most general case, there is a single facet $m$
corresponding to the node $i$ $(q=1)$ and the mesh parameter $k$ has
simply to be set equal to the number of finite elements $e$ on the
interface surface of body 2 converging into the node $i$, divided by
3. This is for instance the case of all the nodes displayed in blue
in Fig.\ref{fig9}. On the other hand, if the orthogonal projection
of the node $i$ onto the interface of body 1 belongs to a boundary
between two facets, then $q=2$ and two node-to-surface interface
elements have to be paired and assembled, one for each facet. In
this case the parameter $k$ is equal to the number of finite
elements on the interface surface of body 2 converging to the node
$i$, divided by $q$ and by 3. This is for instance the case of the
nodes displayed in black in Fig.\ref{fig9} (please refer to the
online version for colors). These operations are summarized in
Algorithm \ref{algo3D}.

\begin{algorithm}[h!]
\caption{3D interface mesh generation algorithm} \label{algo3D}
\vspace*{.1cm}\hrule\vspace*{.1cm} ~~\textbf{Input}: set of nodes
$\IN$ belonging to the interface of body 2 $(\#\IN=n)$; set of
facets $\IM$ $(\#\IM=m)$ of the interface discretization of body 1.
\vspace*{.1cm}\hrule\vspace*{.1cm}
\begin{enumerate}
\item[] \textbf{LOOP} $i=1\dots,n$
\begin{enumerate}
\item Compute the area of a generic facet, $A_{el}$;
\item Determine the number $e$ of facets belonging to the interface of body 2 and converging to the node $i$;
\item Determine the number $q$ of facets of the set $\IM$ with minimum distance
from the node $i$;
\item[] $k\leftarrow e/(3q)$
\item \textbf{LOOP} $j=1,\dots,q$
\item[] Determine the node numbers defining the facet;
\item[] Construct the connectivity matrix of the interface element as a list of the nodes of the facet $j$, plus the node $i$;
\item[] \textbf{END LOOP}
\end{enumerate}
\item[] \textbf{END LOOP}
\end{enumerate}
\vspace*{.1cm}\hrule\vspace*{.1cm} ~~\textbf{Output}: Connectivity
matrix of the node-to-surface interface elements.
\vspace*{.1cm}\hrule\vspace*{.1cm}
\end{algorithm}

\clearpage
\section{Patch tests}

The patch test is performed to show the capability of the proposed
elements to transfer a uniform stress field across an interface in
the case of a uniform loading. Let us consider two blocks of
horizontal side $a=1$ m, and vertical side $b=0.5$ m, made of the
same material and separated by a cohesive interface. Young's modulus
of the bulk is $E=10$ GPa and its Poisson ratio is $\nu=0$. The
lower block is constrained to vertical displacements along its lower
side. One node along the same side is also constrained to horizontal
displacements to avoid rigid body motion. A vertical displacement
$\Delta$ is imposed to the upper side of the second block, in order
to induce Mode I decohesion at the interface whose
traction-separation relation is ruled by the Tvergaard CZM
\cite{TV}. In that CZM, tractions are explicit nonlinear functions
of the relative opening and sliding displacements $g_n$ and $g_t$:
\begin{subequations}\label{TV1}
\begin{align}
\sigma=&\sigma_{\max}\dfrac{g_{n}}{g_{\text{nc}}}P(\lambda)\\
\tau=&\tau_{\max}\dfrac{g_{t}}{g_{\text{tc}}}P(\lambda)
\end{align}
\end{subequations}
where $g_{nc}$ and $g_{tc}$ are the critical opening and sliding
displacements corresponding to complete decohesion under pure Mode I
or Mode II deformation, respectively. The function $P(\lambda)$
reads:
\begin{subequations}\label{TV2}
\begin{align}
P(\lambda)=&\left\{\begin{array}{ll}
             \dfrac{27}{4}\left(1-2\lambda+\lambda^2\right), & \text{for}\, 0\leq\lambda\leq 1 \\
             0, & \text{otherwise}
           \end{array}\right.\\
\lambda=&\sqrt{\left(\dfrac{g_{n}}{g_{nc}}\right)^2+\left(\dfrac{g_{t}}{g_{tc}}\right)^2}
\end{align}
\end{subequations}
For this CZM, the tangent constitutive matrix $\mathbf{C}_P$ has the
following expression:
\begin{equation}\label{TV3}
\begin{aligned}
\mathbf{C}_P=&\left[
             \begin{array}{cc}
\sigma_{\max}\dfrac{P}{g_{nc}}+\sigma_{\max}\dfrac{g_{n}}{g_{nc}}\dfrac{\partial
P}{\partial\lambda}\dfrac{\partial\lambda}{\partial g_{n}} &
             \sigma_{\max}\dfrac{g_{n}}{g_{nc}}\dfrac{\partial P}{\partial\lambda}\dfrac{\partial\lambda}{\partial g_{t}}\\
              \tau_{\max}\dfrac{g_{t}}{g_{tc}}\dfrac{\partial P}{\partial\lambda}\dfrac{\partial\lambda}{\partial
              g_{n}} &
              \tau_{\max}\dfrac{P}{g_{tc}}+\tau_{\max}\dfrac{g_{t}}{g_{tc}}\dfrac{\partial P}{\partial\lambda}\dfrac{\partial\lambda}{\partial
              g_{t}}
             \end{array}
           \right]_P
\end{aligned}
\end{equation}

In 3D, the effective displacement $\lambda$ is computed as follows:
\begin{equation}
\lambda=\sqrt{\left(\dfrac{g_n}{g_{nc}}\right)^2+\left(\dfrac{g_{t1}}{g_{tc}}\right)^2+\left(\dfrac{g_{t2}}{g_{tc}}\right)^2}
\end{equation}
and the cohesive tractions are:
\begin{subequations}\label{tv}
\begin{align}
\sigma=&\sigma_{\max}\dfrac{g_n}{g_{nc}}P(\lambda),\\
\tau_1=&\tau_{\max}\dfrac{g_{t1}}{g_{tc}}P(\lambda),\\
\tau_2=&\tau_{\max}\dfrac{g_{t2}}{g_{tc}}P(\lambda).
\end{align}
\end{subequations}
Similarly, the constitutive matrix $\mathbf{C}_P$ reads:
\begin{equation}
\begin{tiny}
\begin{aligned}
\mathbf{C}_P=&\left[
             \begin{array}{ccc}
\sigma_{\max}\dfrac{P(\lambda)}{g_{nc}}+\sigma_{\max}\dfrac{g_{n}}{g_{nc}}\dfrac{\partial
P(\lambda)}{\partial\lambda}\dfrac{\partial\lambda}{\partial g_{n}}
& \sigma_{\max}\dfrac{g_{n}}{g_{nc}}\dfrac{\partial
P(\lambda)}{\partial\lambda}\dfrac{\partial\lambda}{\partial g_{t1}}
& \sigma_{\max}\dfrac{g_{n}}{g_{nc}}\dfrac{\partial
P(\lambda)}{\partial\lambda}\dfrac{\partial\lambda}{\partial g_{t2}}
\\
\tau_{\max}\dfrac{g_{t1}}{g_{tc}}\dfrac{\partial
P(\lambda)}{\partial\lambda}\dfrac{\partial\lambda}{\partial g_{n}}
&
\tau_{\max}\dfrac{P(\lambda)}{g_{tc}}+\tau_{\max}\dfrac{g_{t1}}{g_{tc}}\dfrac{\partial
P(\lambda)}{\partial\lambda}\dfrac{\partial\lambda}{\partial g_{t1}}
& \tau_{\max}\dfrac{g_{t1}}{g_{tc}}\dfrac{\partial
P(\lambda)}{\partial\lambda}\dfrac{\partial\lambda}{\partial
g_{t2}}\\
\tau_{\max}\dfrac{g_{t2}}{g_{tc}}\dfrac{\partial
P(\lambda)}{\partial\lambda}\dfrac{\partial\lambda}{\partial g_{n}}
& \tau_{\max}\dfrac{g_{t2}}{g_{tc}}\dfrac{\partial
P(\lambda)}{\partial\lambda}\dfrac{\partial\lambda}{\partial
g_{t1}}
&
\tau_{\max}\dfrac{P(\lambda)}{g_{tc}}+\tau_{\max}\dfrac{g_{t2}}{g_{tc}}\dfrac{\partial
P(\lambda)}{\partial\lambda}\dfrac{\partial\lambda}{\partial g_{t2}}
\end{array}
           \right]_P
\end{aligned}
\end{tiny}
\end{equation}

For the tests we select $\sigma_{\max}=10$ Pa, $g_{nc}=g_{tc}=0.3$
m, $\tau_{\max}/\sigma_{\max}=0$. Using these mechanical parameters,
the interface is more compliant than the bulk and we obtain $g_n\sim
\Delta$ at each time step. The corresponding stress field has to be
uniform for equilibrium considerations.

\begin{figure}[h!]
\centering \subfigure[$v$ (4-nodes interface
el.)]{\includegraphics[width=.4\textwidth,angle=0]{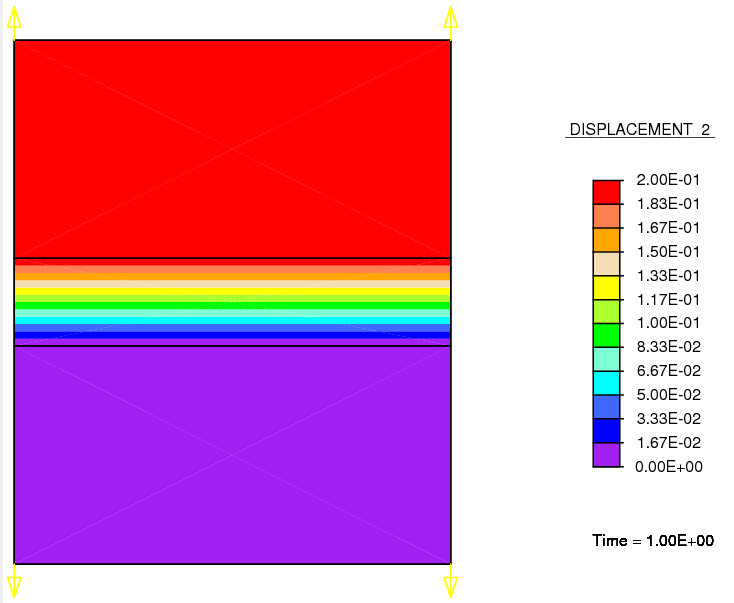}\label{fig10a}}
\subfigure[$\sigma_y$ (4-nodes interface
el.)]{\includegraphics[width=.4\textwidth,angle=0]{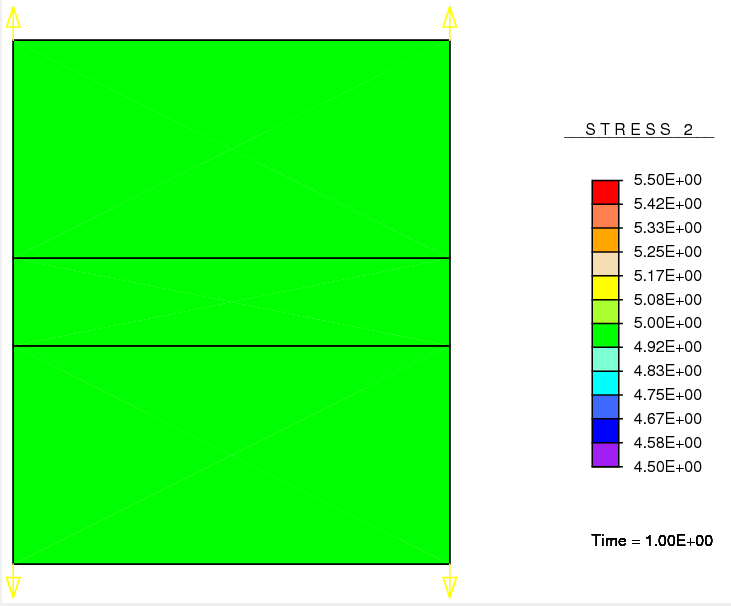}\label{fig10b}}\\
\subfigure[$v$
($h_1/h_2=4$)]{\includegraphics[width=.4\textwidth,angle=0]{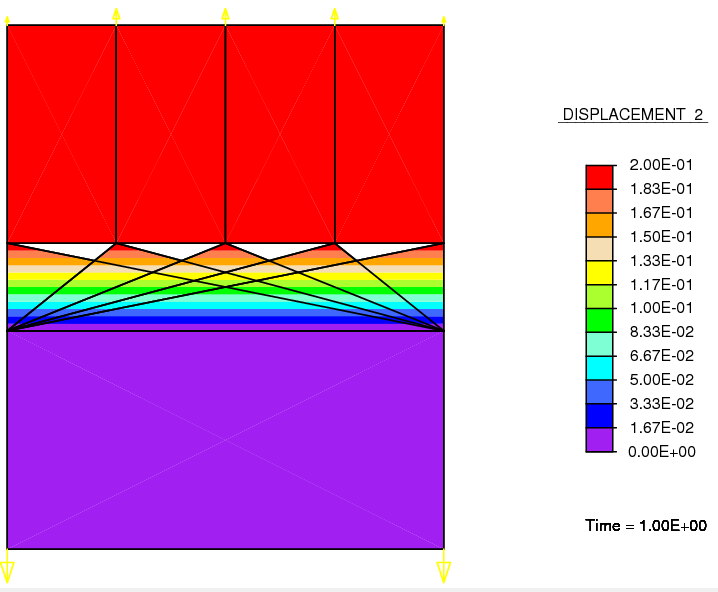}\label{fig10c}}
\subfigure[$\sigma_y$ ($h_1/h_2=4$)]{\includegraphics[width=.4\textwidth,angle=0]{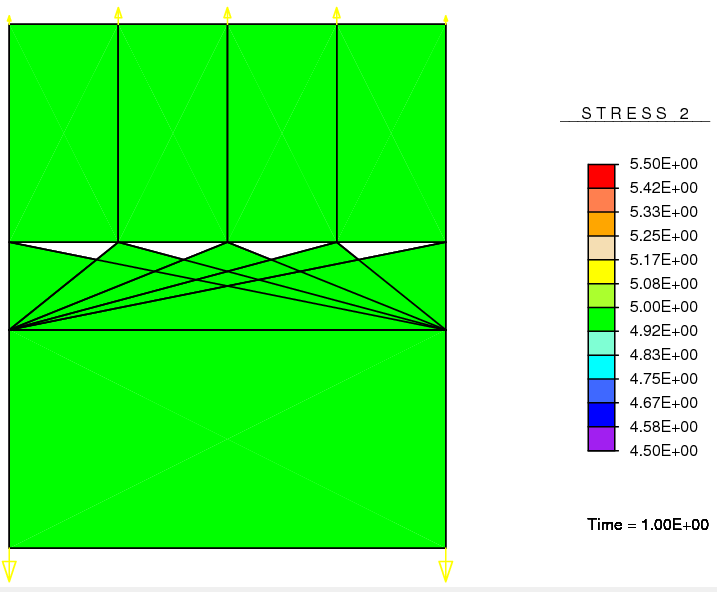}\label{fig10d}}\\
\subfigure[$v$
($h_1/h_2=4$)]{\includegraphics[width=.4\textwidth,angle=0]{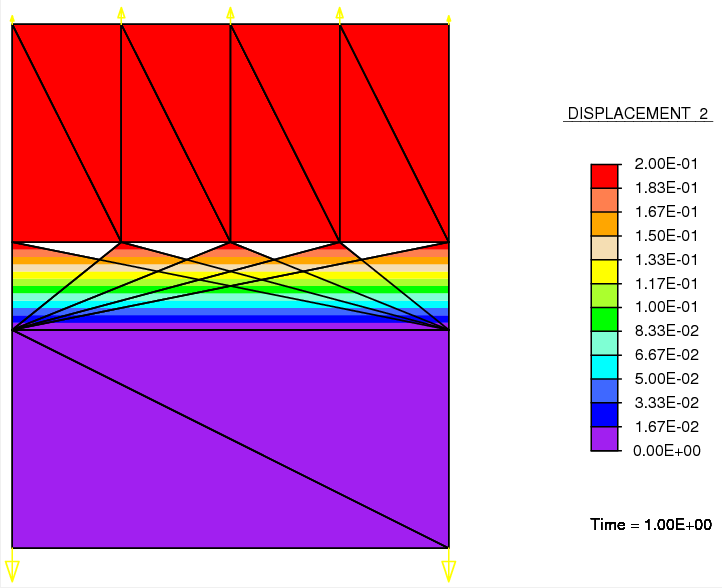}\label{fig10e}}
\subfigure[$\sigma_y$
($h_1/h_2=4$)]{\includegraphics[width=.4\textwidth,angle=0]{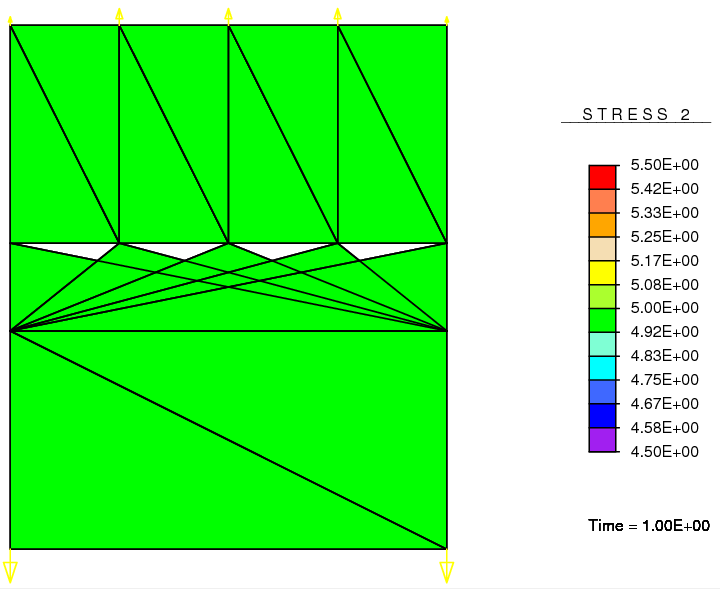}\label{fig10f}}
\subfigure[$v$ (unstructured
mesh)]{\includegraphics[width=.4\textwidth,angle=0]{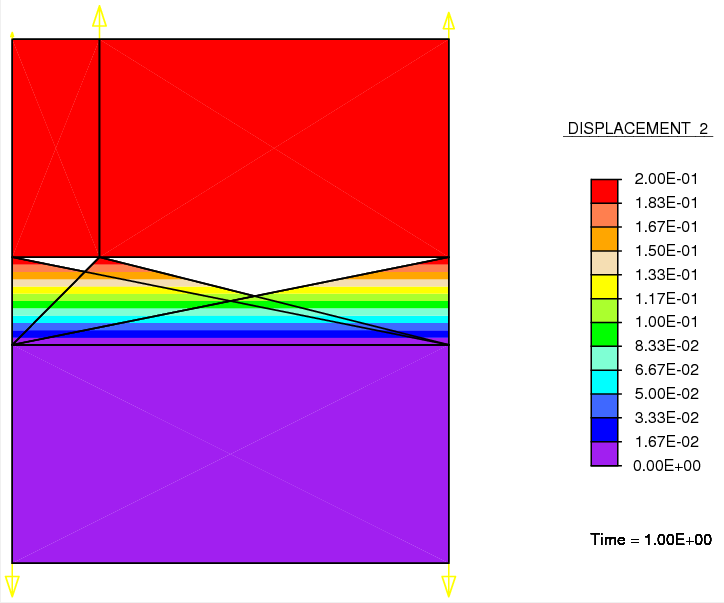}\label{fig10g}}
\subfigure[$\sigma_y$ (unstructured
mesh)]{\includegraphics[width=.4\textwidth,angle=0]{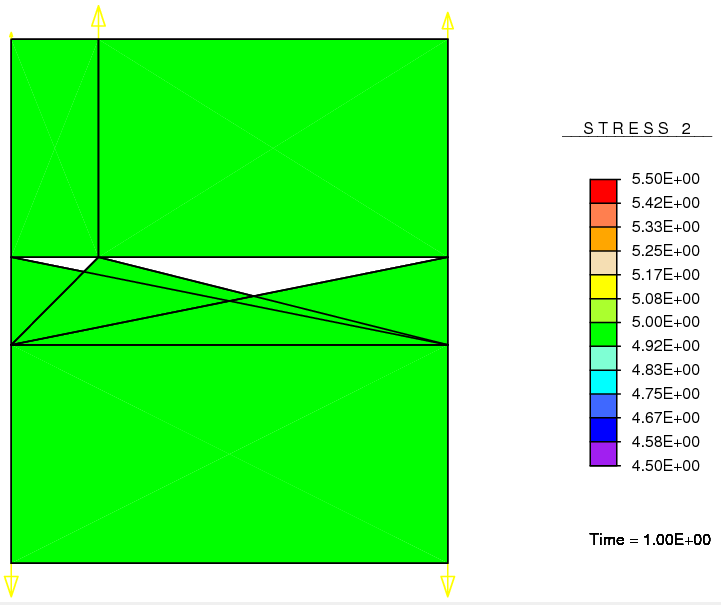}\label{fig10h}}
\caption{Patch test for the 2D node-to-segment interface element:
the uniform stress field is obtained in tension. The benchmark
solution corresponding to a standard 4-nodes interface element is
shown in Figs.\ref{fig10a} and \ref{fig10b}. Figs.\ref{fig10c} and
\ref{fig10d} refer to a discretization with quadrilateral elements
for the continuum; Figs. \ref{fig10e} and \ref{fig10f} refer to
triangular elements; Figs. \ref{fig10g} and \ref{fig10h} correspond
to unstructured finite element meshes.}\label{fig10}
\end{figure}

The contour plots in Fig.\ref{fig10} obtained with $\Delta=0.2$ m
show the vertical displacement field $v$ on the left and the
vertical stress field $\sigma_y$ on the right, for different mesh
discretizations. Figs.\ref{fig10a} and \ref{fig10b} refer to the
classical 4-nodes interface element where matching of nodes at the
interface is required. Figs.\ref{fig10c} and \ref{fig10d} correspond
to different finite element discretizations of the two blocks done
by linear quadrilateral elements, with $h_1/h_2=4$. To avoid the
problem of having unmatched nodes at the interface, node-to-segment
interface elements are used. The uniform stress field is correctly
reproduced as in the case of a standard 4-nodes interface element,
passing the patch test. The same problem is analyzed in
Figs.\ref{fig10e} and \ref{fig10f} with linear triangular elements
used to discretize the continuum. Also in this case, with the same
interface mesh generation procedure as before, the patch test is
passed. Finally, we consider the case of non-structured finite
element meshes, with a nonuniform finite element discretization, see
Figs.\ref{fig10g} and \ref{fig10h}. In this case the length of
influence of each node $l_{el}$ is different from node to node along
the interface, as commented in the previous section.

\begin{figure}[h!]
\centering
\includegraphics[width=.8\textwidth,angle=0]{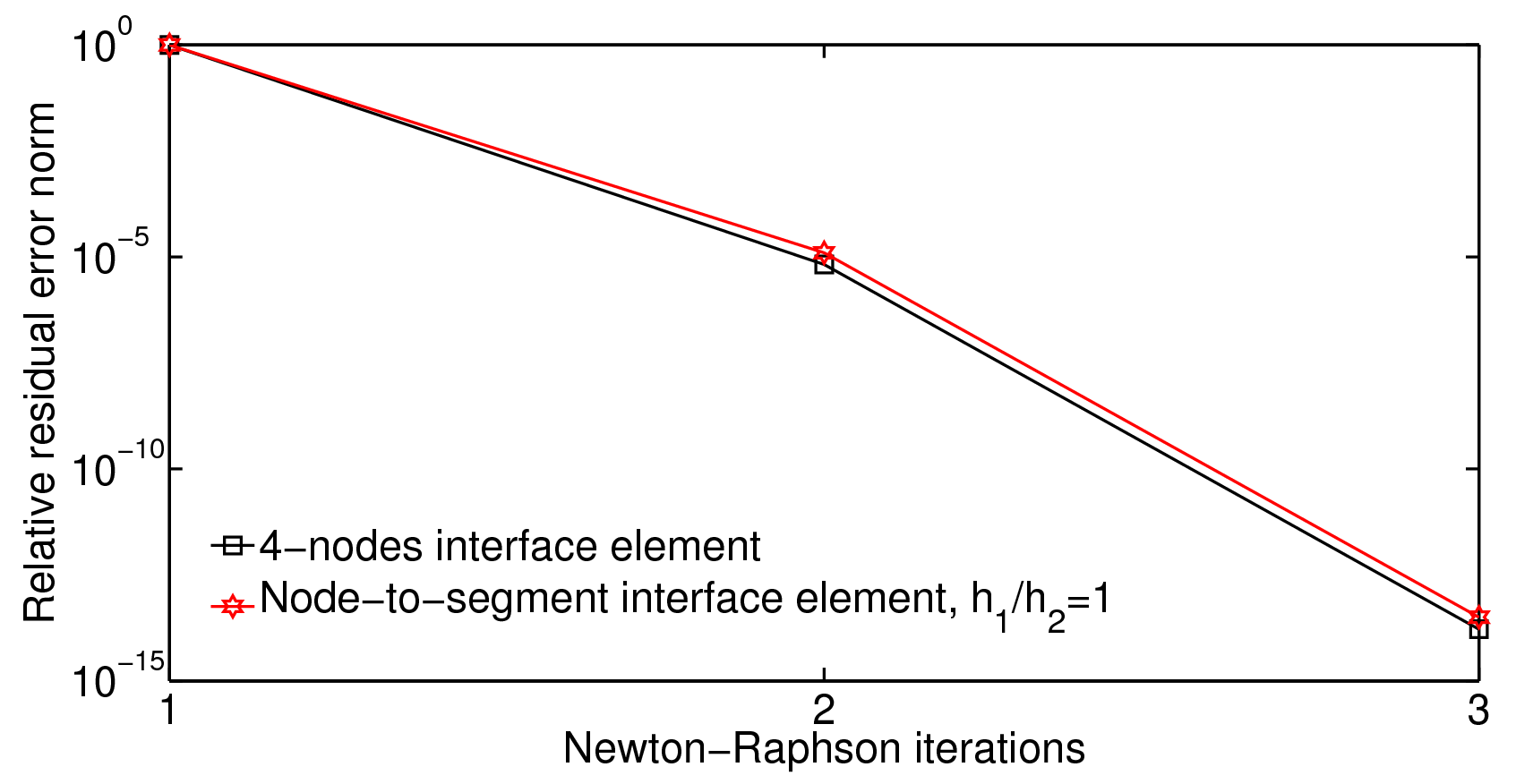}
\caption{Convergence rate of the node-to-segment interface elements
as compared to the standard 4-nodes interface element, for the mesh
discretization of the bulk shown in Fig.\ref{fig10a} and the same
imposed boundary displacements.}\label{fig11}
\end{figure}

The rate of convergence of the Newton-Raphson iterative scheme is
shown in Fig.\ref{fig11} for the 4-nodes interface element (problem
in Fig.\ref{fig10a}) and for the node-to-segment interface element
(problem in Fig.\ref{fig10c}). In both cases quadratic convergence
is achieved.

\begin{figure}[h!]
\centering \subfigure[Standard 8-nodes interface element]{\includegraphics[width=.7\textwidth,angle=0]{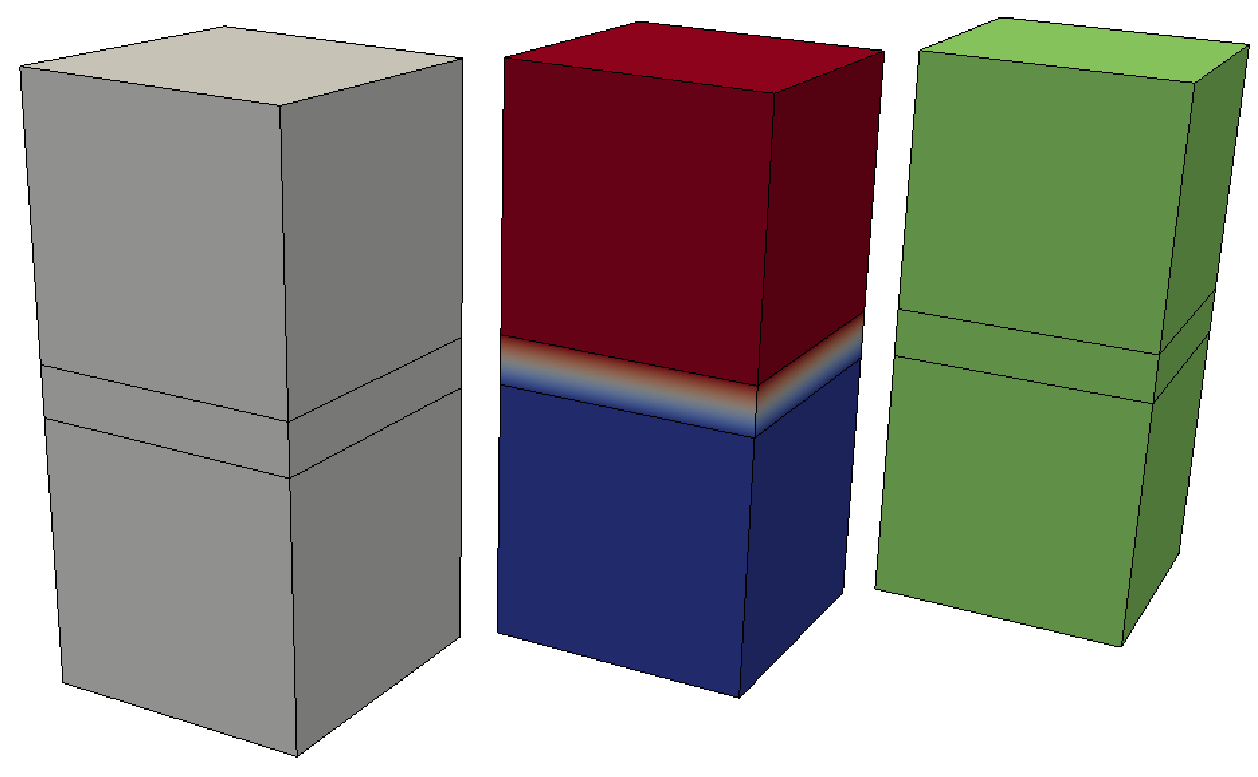}\label{fig12a}}\\
\subfigure[Node-to-surface interface elements,
$h_1/h_2=3$]{\includegraphics[width=.7\textwidth,angle=0]{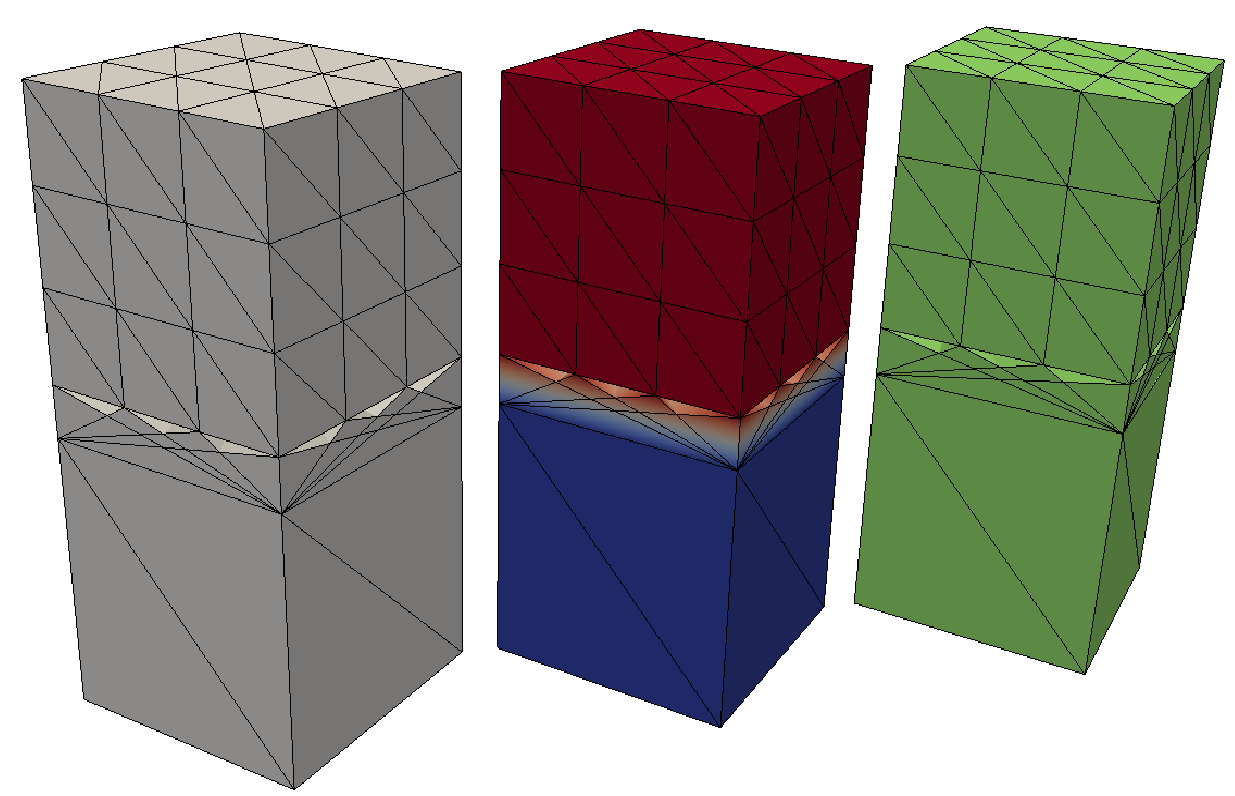}\label{fig12b}}
\caption{Patch test of the 3D node-to-segment interface element: the
uniform stress field is obtained in tension for different
non-matching finite element discretizations of the interface. The
pictures on the left show the finite element discretization, those
in the middle show the displacement field in the $z$-direction, and
those on the right show the vertical stress field in the
$z$-direction.}\label{fig12}
\end{figure}

The patch test is repeated for a 3D problem in Fig.\ref{fig12}.
A standard 8-nodes linear interface element is used in
Fig.\ref{fig12a} in case of matching nodes. The results for the same
uniaxial test problem but with a different discretization of the
blocks realized with linear tetrahedra $(h_1/h_2=3)$ are shown in
Fig.\ref{fig12b}. The algorithm \ref{algo3D} was used to
assemble the interface elements along the interface and the patch
test is passed with a uniform stress field.

\clearpage
\section{Selected applications}

In this section we present selected applications of the new
interface elements to emphasize their usefulness and capabilities in
relation to drawbacks of the standard interface elements pointed out
in the introduction.

\subsection{Peeling of deformable layers from rigid substrates}

\begin{figure}[h!]
\centering \subfigure[Standard 4-nodes interface elements,
$h_1/h_2=1$]{\includegraphics[width=.42\textwidth,angle=0]{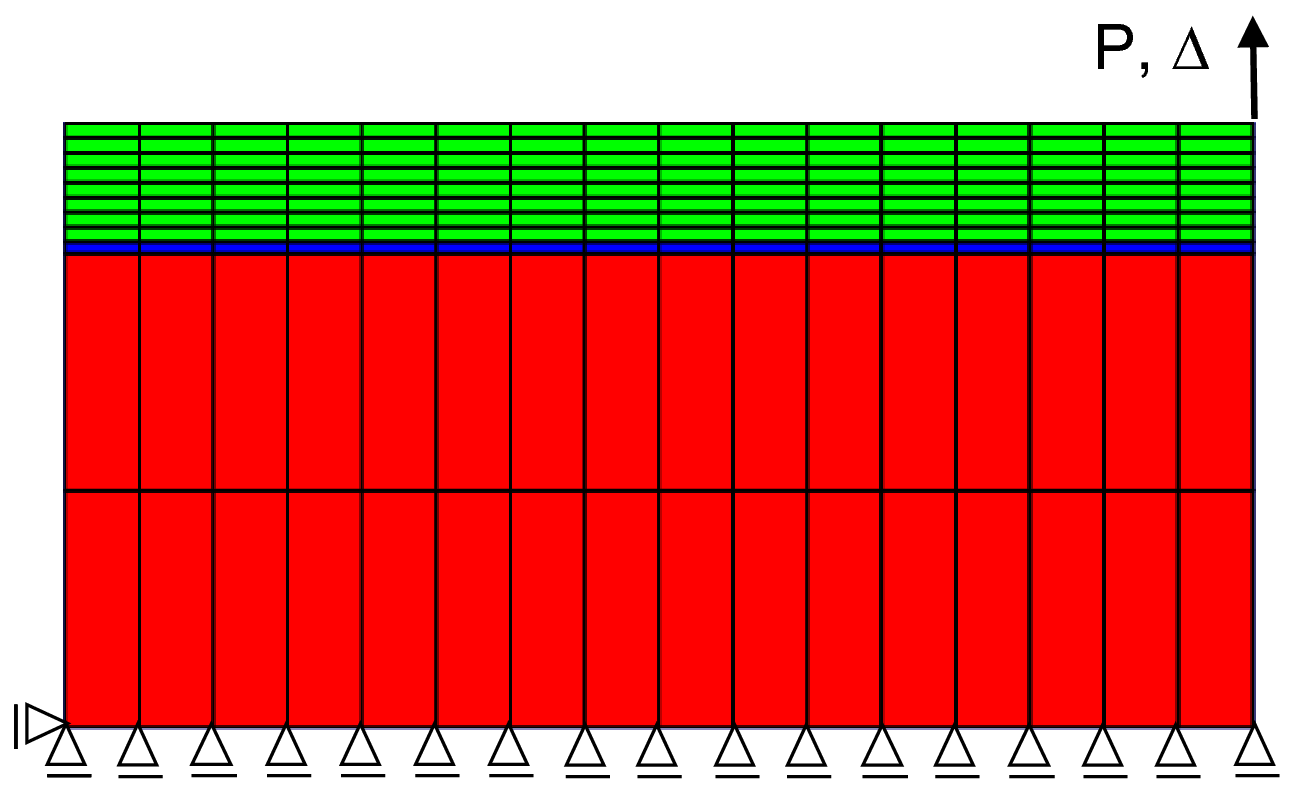}\label{fig13a}}\quad
\subfigure[Node-to-segment interface elements, $h_1/h_2=8$.
]{\includegraphics[width=.42\textwidth,angle=0]{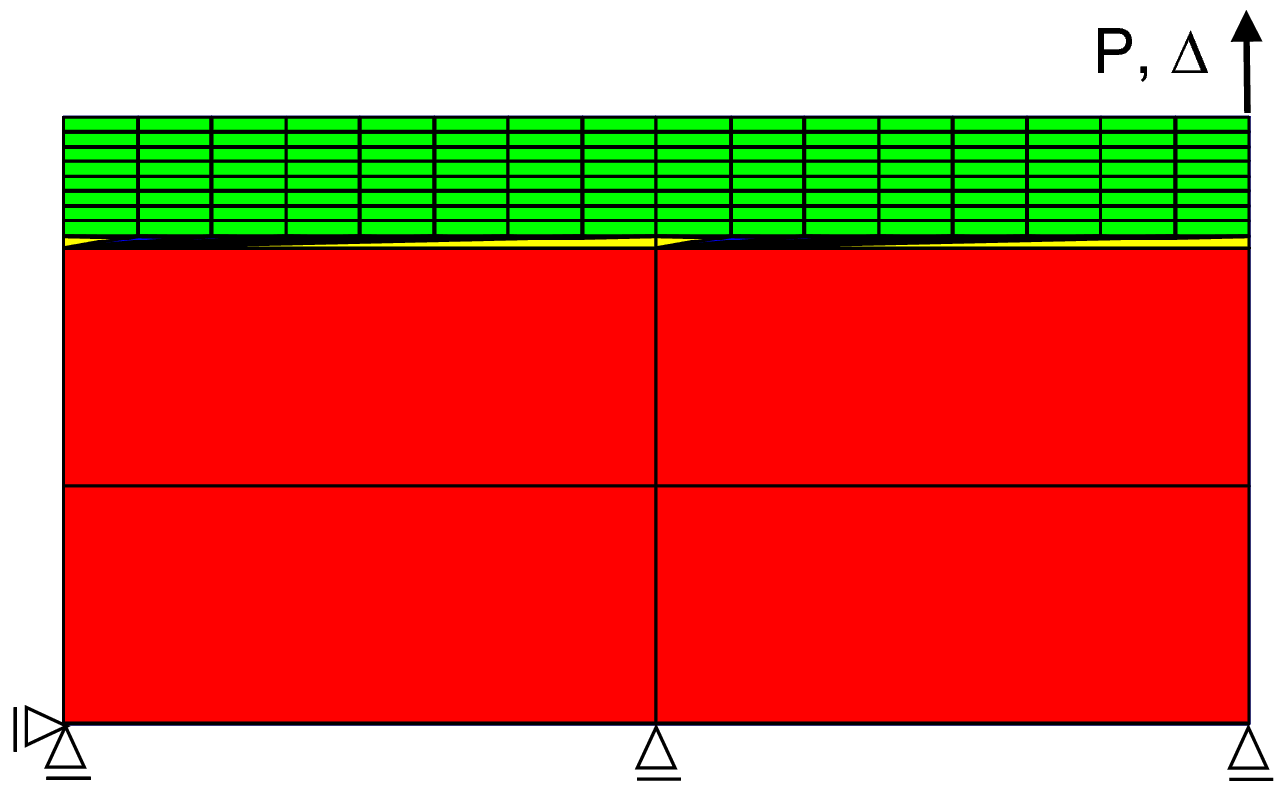}\label{fig13b}}
\caption{Finite element discretizations for the peeling
test.}\label{fig13}
\end{figure}

Let's consider the problem of peeling of a deformable layer from a
stiff substrate, see Fig.\ref{fig13}. The numerical simulation of
this problem requires a fine discretization of the deformable layer
through its width to capture its bending deformation. The interface
needs a fine discretization as well, to correctly resolve the
cohesive tractions. On the other hand, a much coarser discretization
for the substrate would be adequate, since it behaves almost as a
rigid body. This is therefore a typical problem that could be
efficiently handled by using two different mesh discretizations at
the interface. An example of a finite element discretization by
using standard 4-nodes interface elements, requiring matching of the
nodes of the finite elements of body 1 and of body 2 at the
interface, is shown in Fig.\ref{fig13a}. With the use of the
node-to-segment interface elements we can keep fine the
discretization of the deformable layer and we can coarsen the
discretization of the substrate, see Fig.\ref{fig13b} with
$h_1/h_2=8$. The substrate is restrained along its base and a
vertical displacement $\Delta$, linearly increasing with time, is
imposed to the layer as shown in Fig.\ref{fig13}. The peeling force
$P$ is computed as a reaction force for each value of $\Delta$.

The Young's modulus of the layer is $E_2=1\times 10^5$ Pa and that
of the substrate is $E_1=1\times 10^{10}$ Pa. In spite of the
simplicity of the test, the significant mismatch in the elastic
properties and the nonuniform traction distribution along the
interface during peeling provides a critical benchmark problem for
testing the accuracy of the new element in predicting the global
force-displacement curve and the local traction distribution along
the interface.

The horizontal size of the blocks is $L=1$ m, and the widths of the
substrate and of the layer are set equal to 0.4 m and 0.1 m,
respectively. We consider the cohesive zone model by Tveergard
\cite{TV} as detailed in the previous section, considering two
cases: $(i)$ $\sigma_{\max}=30$ Pa, $g_{nc}=g_{tc}=0.1$ m,
$\tau_{\max}/\sigma_{\max}=0$; $(ii)$ $\sigma_{\max}=30$ Pa,
$g_{nc}=g_{tc}=0.1$ m, $\tau_{\max}/\sigma_{\max}=1$.

\begin{figure}[h!]
\centering
\includegraphics[width=.85\textwidth,angle=0]{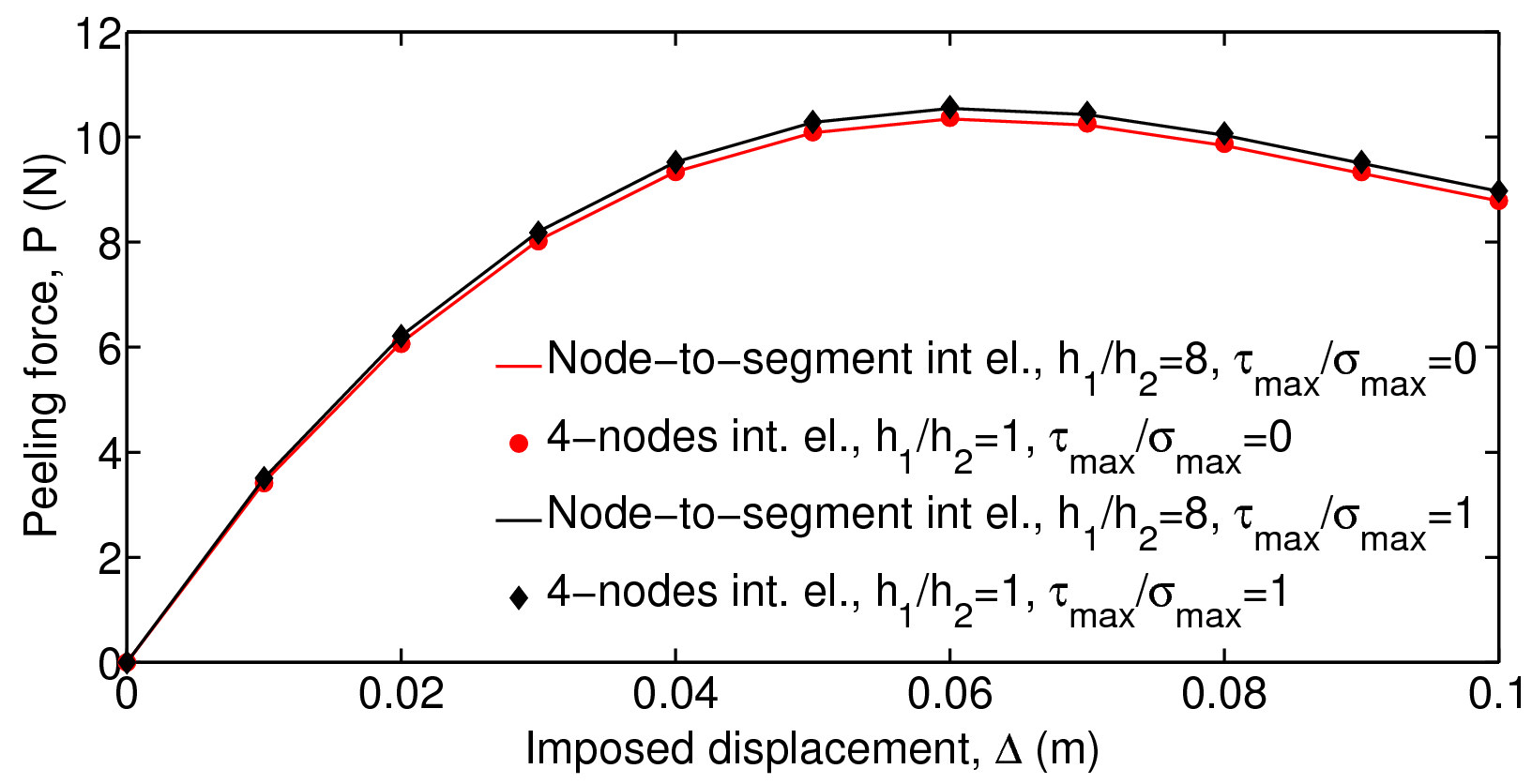}
\caption{Force-displacement curves for the peeling test in
Fig.\ref{fig13}, for two different values of
$\tau_{\max}/\sigma_{\max}$. Benchmark results obtained by using
4-nodes interface elements are shown with bullets.}\label{fig14}
\end{figure}

The peeling force vs. imposed displacement curves for the two cases
and for the two different mesh discretizations in Fig.\ref{fig13}
are shown in Fig.\ref{fig14}. The solution based on the
node-to-segment interface elements is the same as that for the
classic 4-nodes interface element. However, note that the new
formulation leads to a significant reduction of finite elements used
to discretize the rigid block. In terms of computation time, in
spite of the relative simplicity of this test problem, a gain of
$18\%$ is achieved by using the discretization in Fig.\ref{fig13b}
with respect to that in Fig.\ref{fig13a}.

\begin{figure}[h!]
\centering
\subfigure[Dimensionless separation, $\lambda$]{\includegraphics[width=.75\textwidth,angle=0]{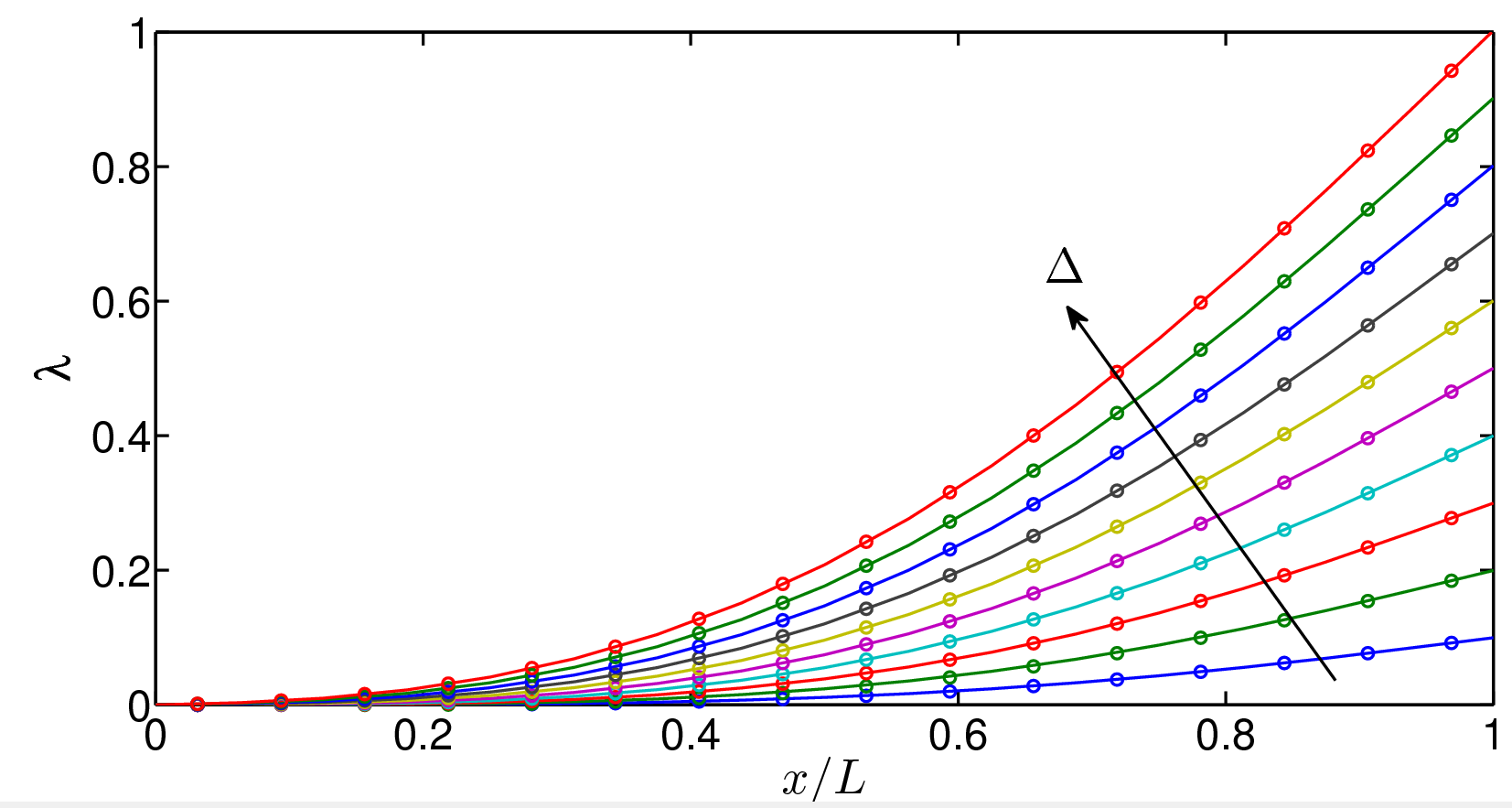}\label{fig15a}}\\
\subfigure[Normal cohesive traction, $\sigma$]{\includegraphics[width=.75\textwidth,angle=0]{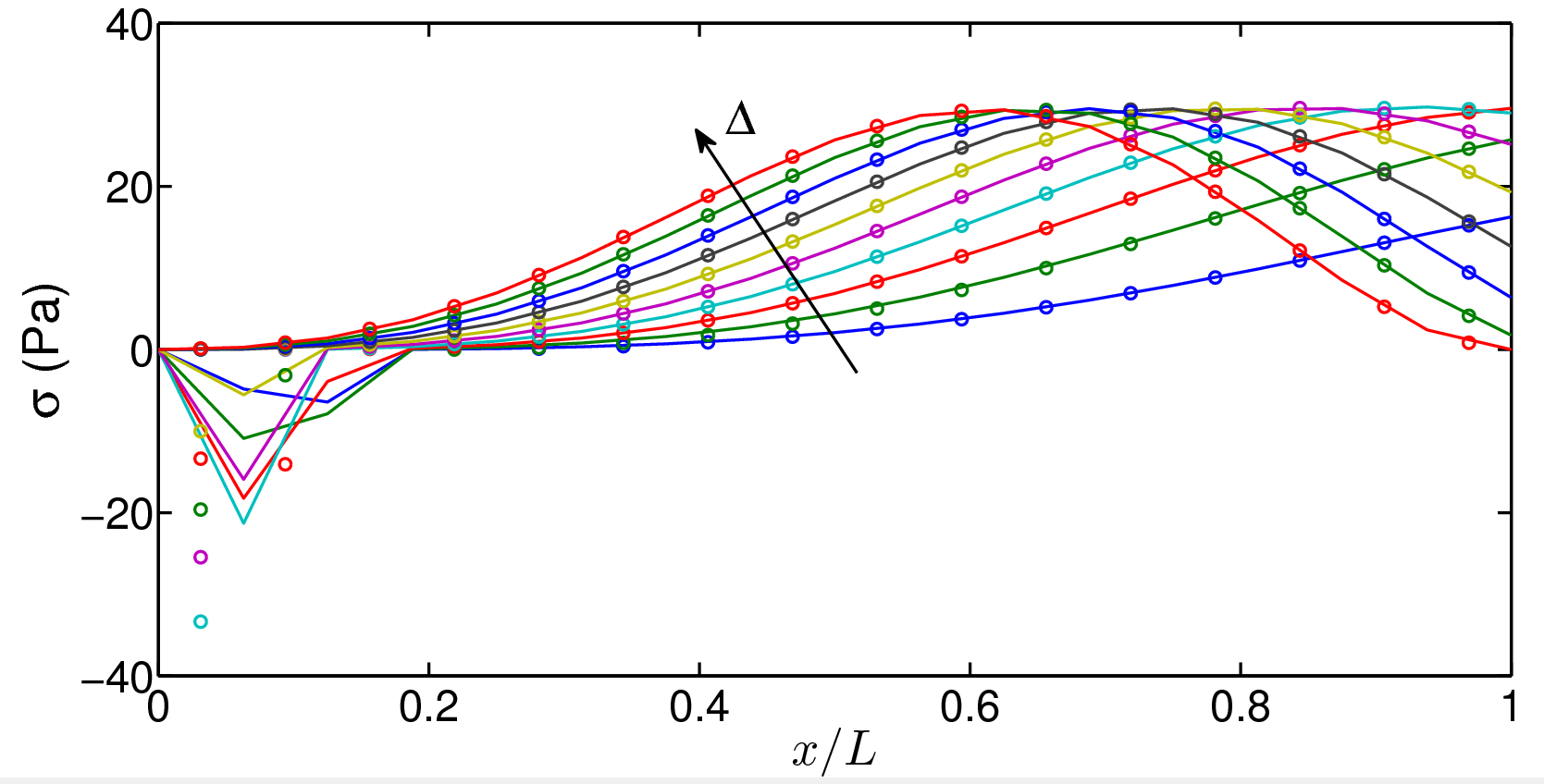}\label{fig15b}}\\
\subfigure[Tangential cohesive traction,
$\tau$]{\includegraphics[width=.75\textwidth,angle=0]{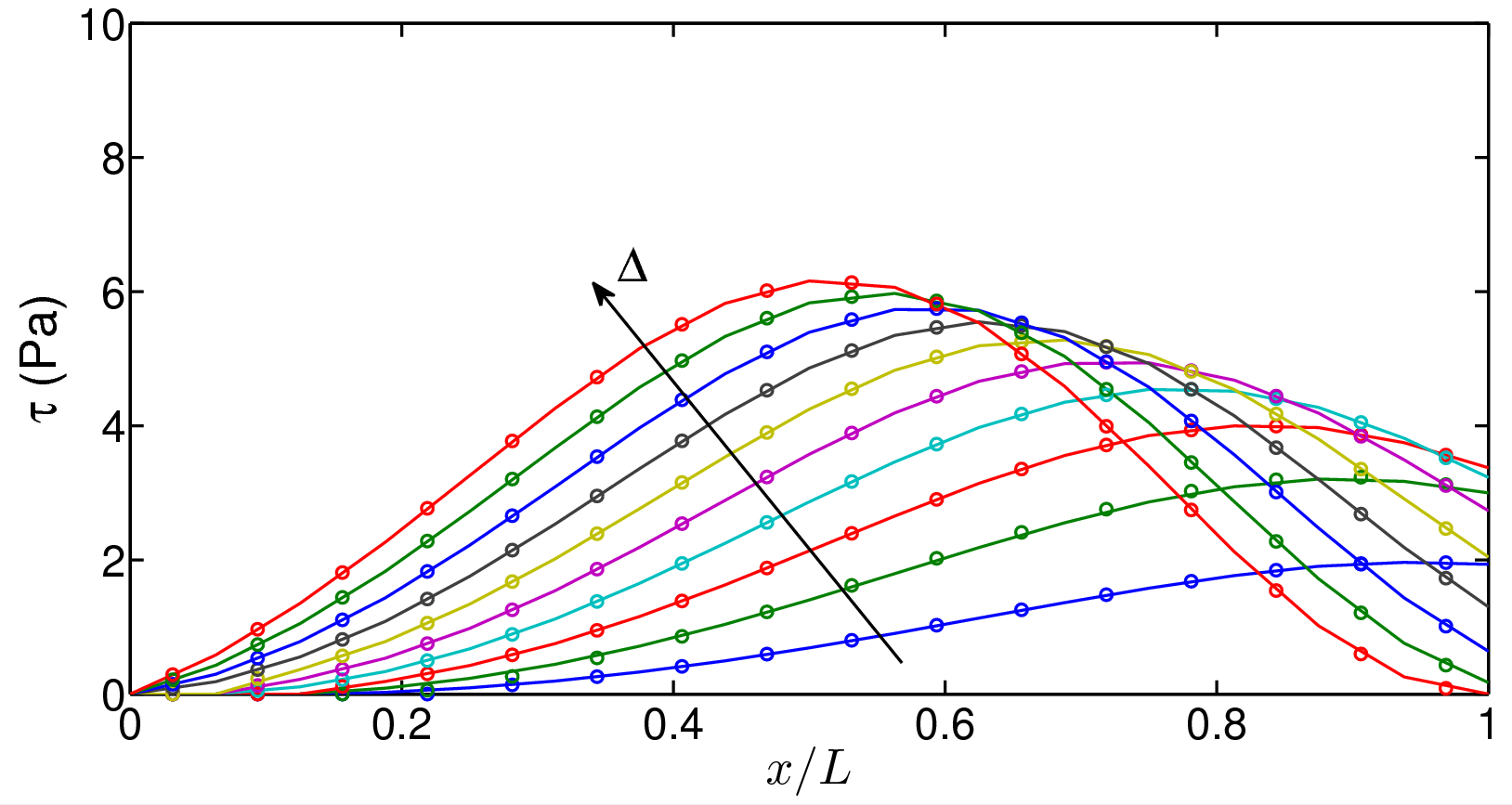}\label{fig15c}}
\caption{Local response of the proposed interface element (solid
lines) vs. standard interface element (dots) for different far-field
imposed displacements $\Delta$ in Fig. \ref{fig14}.}\label{fig15}
\end{figure}

The quality of the local solution can be assessed by examining the
predicted effective dimensionless crack separation $\lambda$ and the
normal and tangential cohesive tractions vs. $x/L$, where the
variable $x$ defines the position along the interface ranging from
$x/L=0$ at the left hand side up to $x/L=1$ at the right hand side.
The results based on the node-to-segment interface element
discretization are shown in Fig.\ref{fig15} for the previous test
problem with $\tau_{\max}/\sigma_{\max}=1$, while the results based
on the standard interface element discretization are shown with
dots. The agreement between the two methods is excellent for all the
interface quantities in the opening regime, i.e., for $g_n>0$. For
the portion of the interface in contact near $x/L=0$, on the other
hand, a difference in the distributions of the normal traction
$\sigma$ is noticed. In the present framework, for both finite
element discretization schemes, the contact constraint was enforced
by the penalty method with a penalty parameter $\epsilon=1\times
10^{18}$ Pa/m. A possible improvement of the present work is
expected by incorporating more accurate contact techniques like the
\emph{mortar method} \cite{M1,M2,M3,M4} in the interface element
formulation, splitting the treatment of the contact stage from the
debonding one.

\clearpage

\subsection{Modelling cohesive interfaces with non propagating crack tips}

\begin{figure}[h!]
\centering \subfigure[]{\includegraphics[width=.4\textwidth,angle=0]{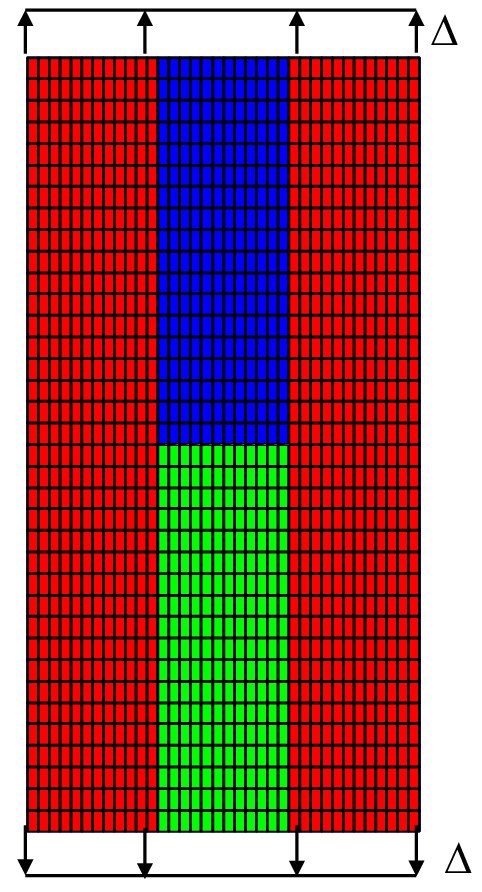}\label{fig16a}}\\
\subfigure[]{\includegraphics[width=.45\textwidth,angle=0]{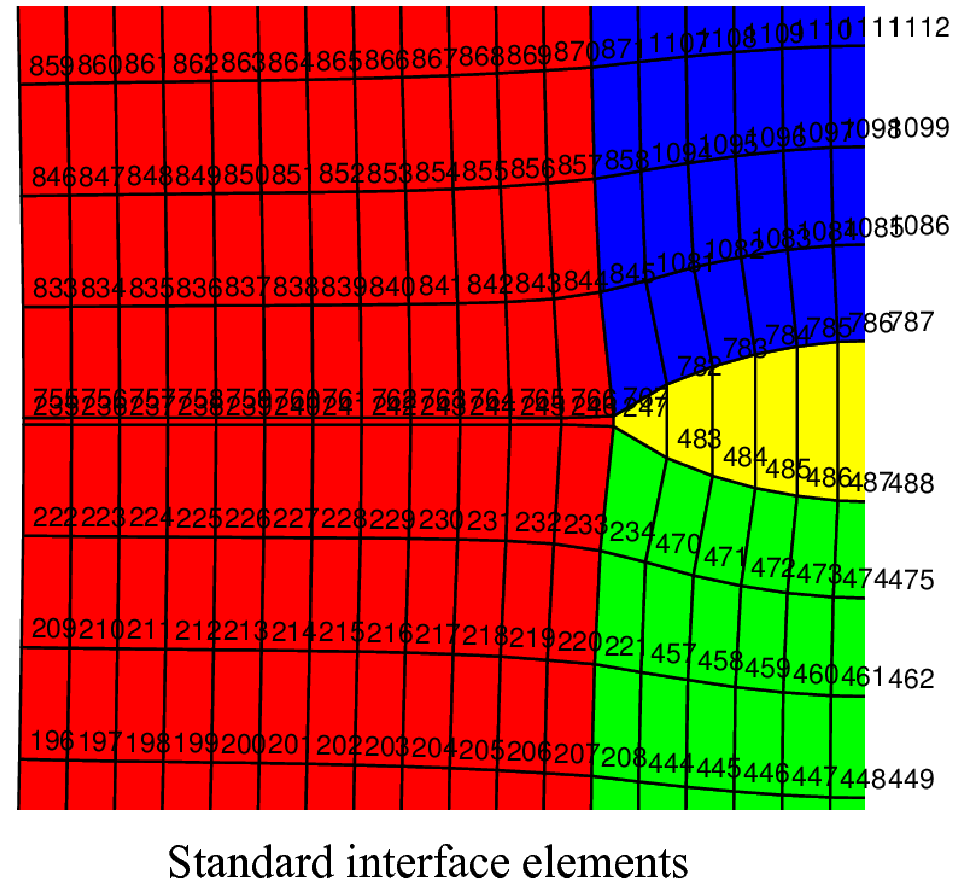}\label{fig16b}}
\subfigure[]{\includegraphics[width=.45\textwidth,angle=0]{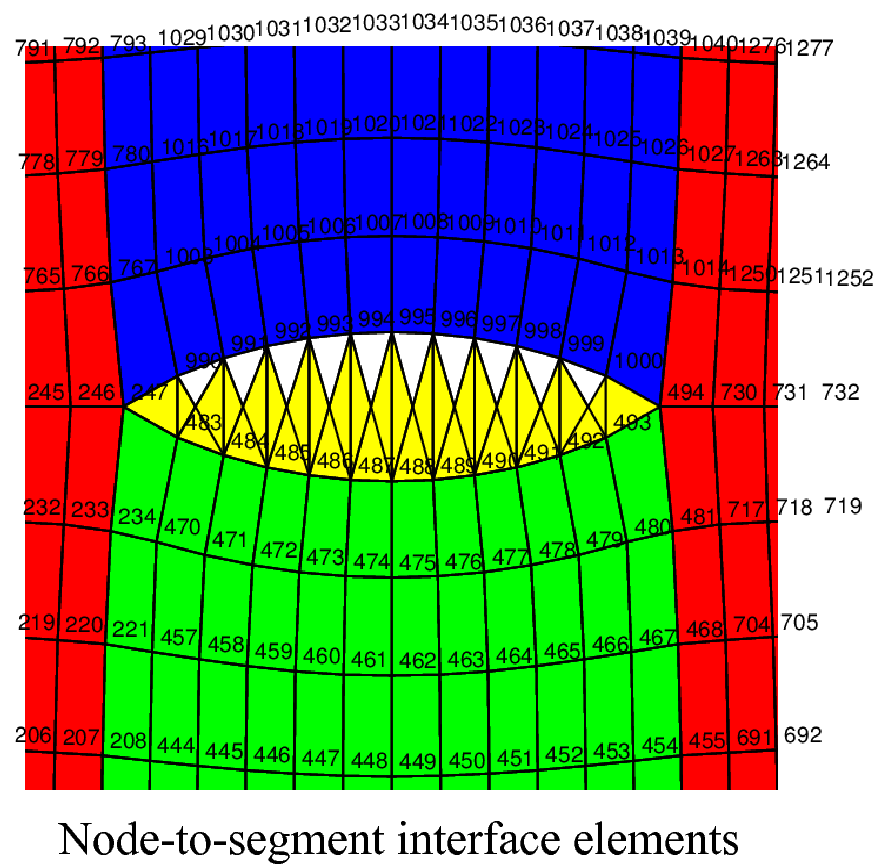}\label{fig16c}}
\caption{A central cohesive crack with non-propagating crack tips
inserted into a composite elastic plate in tension.}\label{fig16}
\end{figure}

Another problem where the new interface elements can be useful
concerns modelling cohesive interfaces whose tips are expected not
to propagate. In reference to Fig.\ref{fig16a}, this can be the case
of an interface between two strips shown in blue and green of 6 m
height and 1 m width (please refer to the online version of this
article for colors), fully bonded along vertical interfaces to two
continuous external strips also 1 m width, shown in red. Assuming
that the cohesive interface cannot propagate into the external
layers and it cannot lead to debonding of the vertical interfaces,
interface elements have to be inserted along the crack line only.

Apart from the possibility of using different mesh discretizations
for the blue and the green strips dictated by error control
procedures, treatment of the crack tips is also a problem to be deal
with. In fact, if we use standard 4-nodes interface elements, the
crack tip nodes have to be duplicated as well and, as a consequence,
thin solid elements have to be inserted along a line inside the red
layers till reaching the external boundaries, see the discretization
in the bulk of the red layer on the left of the crack tip in
Fig.\ref{fig16b}. The same takes place ahead of the opposite crack
tip. The width of these elements cannot be set equal to zero, but
has to be taken as small as possible to model a sharp crack tip. The
elongated shape of the continuum elements ahead of the crack tip is
of course not ideal as far as their accuracy is concerned.

This problem can be completely avoided without affecting the way the
finite element discretization of the red strips is performed by
using node-to-segment interface elements to deal with the crack tip
dicretization. The regions corresponding to the different material
domains can be meshed independently from each other and their common
nodes tied everywhere, except between the blue and the green
regions. Zero-thickness node-to-segment interface elements are then
inserted along this interface by applying the Algorithm
\ref{algo2D}. Alternatively, classical interface elements can be
used at this stage for the cohesive description of the crack. At the
crack tip, on the other hand, the node-to-segment interface element
able to work with three nodes, namely nodes 247 and 483 in
Fig.\ref{fig16c} (defining the segment) and the node 998 for the
crack tip on the left, is used. Similarly, for the opposite crack
tip, the nodes 493 and 494 (defining the segment) and the node 1000
for the crack tip are used to generate the other node-to-segment
interface element. This avoids the duplication of the nodes 257 and
494 and hence the finite element discretization of the continuous
red stripes is no longer affected by the presence of the crack.

\begin{figure}[h!]
\centering
\includegraphics[width=.9\textwidth,angle=0]{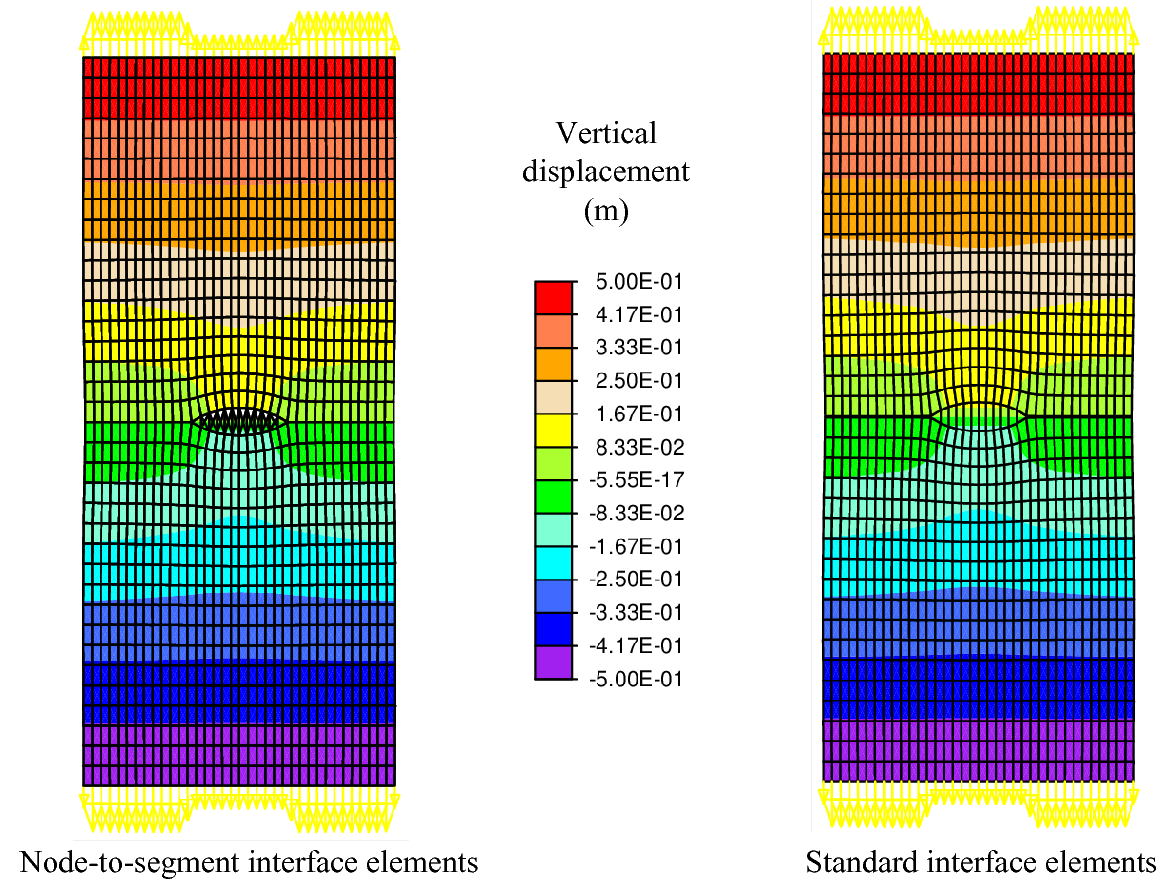}
\caption{Contour plots of the vertical displacement field related to
the problem in Fig.\ref{fig16}, discretized with node-to-segment
interface elements or with standard interface
elements.}\label{fig17}
\end{figure}

The contour plots of the vertical displacement field are shown in
Fig.\ref{fig17} for the sake of completeness, considering a Young's
modulus of the red strips equal to $2\times 10^5$ Pa, a Young's
modulus of the blue and green strips equal to $1\times 10^5$ Pa,
vanishing Poisson's ratios, CZM parameters $\sigma_{\max}=1000$ Pa,
$\tau_{\max}/\sigma_{\max}=0$, $g_n=g_t=0.3$, and $\Delta=0.5$ m.
The crack opening profile is exactly the same in the two cases.

\clearpage

\section{Conclusions and perspectives}

Two novel interface elements have been proposed to discretize
interfaces with non-matching nodes. The formulation is based on a particularization of
node-to-segment and node-to-surface contact elements used in contact
mechanics under the assumption of small displacements.

Mesh generation algorithms for 2D and 3D problems have also been
presented, with a significant simplification over the classical
meshing procedure required in the case of standard 2D and 3D
interface elements. The proposed formulation allows the generation
of finite element meshes of different bodies/material regions
separately and depending on the needs, without the constraint of
imposing the same finite element discretization at the interface.
Therefore, the cumbersome procedure of meshing domains as fully
bonded and then duplicating nodes at the interface can be completely
avoided. This is particularly useful in the case of polygonal
domains separated by cohesive interfaces, as in 2D and 3D
polycrystalline materials, and in delamination problems where one
material component is much stiffer than the other one, as in
fiber-reinforced composites or in peeling tests of tapes from stiff
substrates.

In terms of computational efficiency, uncoupling the mesh
discretization of material regions allows a better control of mesh
quality in computational materials science applications, and a
selective reduction of number of finite elements in the case of
fracture problems involving bodies with very different elastic
properties. In the 2D peeling test, for instance, a saving of $18\%$
in computation time was achieved in spite of its coarse
discretization. The local solution was also found quite accurate
with respect to the standard interface element discretization,
especially in terms of equivalent dimensionless gap and cohesive
tractions along the portion of the interface experiencing debonding.
In the part of the interface in contact, on the other hand, a
discrepancy in the predictions has been observed. This suggests
further developments of the proposed formulations by splitting the
treatment of the debonding stage from the contact one. In the case
of a negative crack opening (contact), the mortar method using dual
spaces for the Lagrange multiplier \cite{M1} could be exploited. The
mortar method introduces the continuity condition at interfaces in
integral (global) form, rather than as nodal (local) constraints,
with an improvement in the description of the contact predictions.
This route is left for further developments and opens the
possibility to create a new prototype of interface element able to
deal with non-matching nodes during delamination and contact,
particularly useful to accurately simulate cyclic loading conditions
with alternating tension and compression stress states.

Finally, the second example has shown that the node-to-segment and
the node-to-surface finite elements can used to discretize interface
cracks with non-propagating crack tips. If employed to discretize
the crack tips, then the finite element discretization of the
neighboring material domains can be left completely unmodified,
without the need of introducing continuum finite elements with very
elongated shapes or changing the finite element discretization of
the strips by using non uniform triangular meshes.

\vspace{1em} \addcontentsline{toc}{section}{Acknowledgements}
\noindent\textbf{Acknowledgements} \vspace{1em}

\noindent The research leading to these results has received funding
from the European Research Council under the European Union's
Seventh Framework Programme (FP/2007-2013) / ERC Grant Agreement n.
306622 (ERC Starting Grant ``Multi-field and multi-scale
Computational Approach to Design and Durability of PhotoVoltaic
Modules" - CA2PVM; PI: Prof. M. Paggi).

\section*{Appendix: coefficients entering the rotation matrices}

For the node-to-segment interface element, the components of the
normal and tangential unit vectors entering the definition of the
rotation matrix $\mathbf{R}$ are related to the coordinates of the
nodes 1 and 2, $\mathbf{x}_i=(x_i,y_i)^{\mathrm{T}}$ $(i=1,2)$, see
Fig.\ref{fig6b}. Their expression is the following:
\begin{subequations}
\begin{align}
n_x&=-\dfrac{y_2-y_1}{\|\mathbf{x}_2-\mathbf{x}_1\|}\\
n_y&=\dfrac{x_2-x_1}{\|\mathbf{x}_2-\mathbf{x}_1\|}\\
t_x&=\dfrac{x_2-x_1}{\|\mathbf{x}_2-\mathbf{x}_1\|}\\
t_y&=\dfrac{y_2-y_1}{\|\mathbf{x}_2-\mathbf{x}_1\|}
\end{align}
\end{subequations}

For the node-to-surface interface element, on the other hand, the
definition of the coefficients entering the rotation matrix requires
to preliminary introduce the vectors $\mathbf{x}_{21}$ and
$\mathbf{x}_{31}$ of components (see Fig. \ref{fig7b}):
\begin{subequations}
\begin{align*}
x_{21}&=x_2-x_1,\quad y_{21}=y_2-y_1,\quad z_{21}=z_2-z_1\\
x_{31}&=x_3-x_1,\quad y_{31}=y_3-y_1,\quad z_{31}=z_3-z_1
\end{align*}
\end{subequations}

After computing the area $A$ of the facet $1-2-3$:
\begin{equation}
A=\dfrac{\sqrt{(y_{21}z_{31}-z_{21}y_{31})^2+(z_{21}x_{31}-x_{21}z_{31})^2+(z_{21}y_{31}-y_{21}z_{31})^2}}{2}
\end{equation}
the components of the normal and tangential unit vectors are finally
provided by the following relations:
\begin{subequations}
\begin{align}
n_x&=\dfrac{y_{21}z_{31}-z_{21}y_{31}}{2A}\\
n_y&=\dfrac{z_{21}x_{31}-x_{21}z_{31}}{2A}\\
n_z&=\dfrac{z_{21}y_{31}-y_{21}z_{31}}{2A}\\
t_{1x}&=\dfrac{x_{21}}{\sqrt{x_{21}^2+y_{21}^2+z_{21}^2}}\\
t_{1y}&=\dfrac{y_{21}}{\sqrt{x_{21}^2+y_{21}^2+z_{21}^2}}\\
t_{1z}&=\dfrac{z_{21}}{\sqrt{x_{21}^2+y_{21}^2+z_{21}^2}}\\
t_{2x}&=t_{1x}n_{y}-t_{1y}n_{z}\\
t_{2y}&=t_{1x}n_{z}-t_{1z}n_{x}\\
t_{2z}&=t_{1y}n_{x}-t_{1x}n_{y}
\end{align}
\end{subequations}

\section*{References}

\bibliographystyle{elsarticle-num}

\end{document}